\documentclass[12pt,preprint,apjfonts,apj]{aastex}
\usepackage{graphicx,amsmath,amssymb}
\usepackage{longtable,lscape}  
\usepackage{hyperref}

\begin{document}

\title{VLBI observations of 10 CSO candidates: expansion velocities of hot spots}
\shortauthors{An et al.}
\shorttitle{Expansion velocities of 10 CSO candidates}
\date{\today}

\author{
Tao~An\altaffilmark{1,2,3},
Fang~Wu\altaffilmark{1,3,4}, 
Jun~Yang\altaffilmark{5}, 
Gregory B.~Taylor\altaffilmark{6}, 
Xiaoyu~Hong\altaffilmark{1,3},  \\
Willem A.~Baan\altaffilmark{2}, 
Xiang~Liu\altaffilmark{7,3},
Min~Wang\altaffilmark{8},
Hongbo~Zhang\altaffilmark{9},
Weihua~Wang\altaffilmark{1,3},\\
Xi~Chen\altaffilmark{1,3},
Lang~Cui\altaffilmark{7}, 
Longfei~Hao\altaffilmark{8}, 
Xinying~Zhu\altaffilmark{9}
}
\altaffiltext{1}{Shanghai Astronomical Observatory, CAS, 200030, Shanghai, China; antao@shao.ac.cn}
\altaffiltext{2}{ASTRON, PO Box 2, 7990 AA Dwingeloo, The Netherlands; an@astron.nl}
\altaffiltext{3}{Key Laboratory of Radio Astronomy, Chinese Academy of Sciences, China}
\altaffiltext{4}{Graduate University of the Chinese Academy of Sciences, 100049, Beijing, China}
\altaffiltext{5}{Joint Institute for VLBI in Europe, Postbus 2, 7990AA Dwingeloo, The Netherlands}
\altaffiltext{6}{Department of Physics and Astronomy, University of New Maxico, Albuquerque NM, 87131, USA}
\altaffiltext{7}{National Astronomical Observatories/Xinjiang Astronomical Observatory, CAS, Urumqi 830011, China}
\altaffiltext{8}{National Astronomical Observatories/Yunnan Observatory, CAS, Kunming 650011, China}
\altaffiltext{9}{National Astronomical Observatories, CAS, Beijing 100012, China}

\begin{abstract}
Observations of ten Compact Symmetric Objects ({\rm CSO}) candidates have been made with the Very Long Baseline Array at 8.4 GHz in 2005 and with a combined Chinese and European VLBI array at 8.4 GHz in 2009. The 2009 observations incorporate for the first time the two new Chinese telescopes at Miyun and Kunming for international astrophysical observations. The observational data, in combination with archival VLBA data from previous epochs, have been used to derive the proper motions of the VLBI components.   Because of the long time baseline of $\sim$16 years of the VLBI data sets, the expansion velocities of the hot spots can be measured at an accuracy as high as $\sim$1.3 $\mu$as yr$^{-1}$. Six of the ten sources are identified as CSOs with a typical double or triple morphology on the basis of both spectral index maps and their mirror-symmetry of proper motions of the terminal hot spots. The compact double source J1324+4048 is also identified as a CSO candidate. Among the three remaining sources, J1756+5748 and J2312+3847 are identified as core-jet sources with proper motions of their jet components relating to systemic source expansion. The third source J0017+5312 is likely also a core-jet source, but a robust detection of a core is needed for an unambiguous identification. The kinematic ages of the CSOs derived from proper motions range from 300 to 2500 years. The kinematic age distribution of the CSOs confirm an overabundance of compact young CSOs with ages less than 500 years. CSOs with known kinematic ages may be used to study the dynamical evolution of extragalactic radio sources at early stages. 
 
\end{abstract}


\section{INTRODUCTION}

Compact Symmetric Objects (CSOs) are a class of extragalactic radio sources that have sub-galactic ($\lesssim$1 kpc) sizes and are characterized by compact symmetric double components. They were first found in the VLBI observations made by Phillips \& Mutel (1980, 1982), and the term CSO was introduced by Wilkinson et al. (1994). The compact double components  show steep radio spectra, similar to the lobes in large radio galaxies, and are extended regions of radio emitting plasma expanding into the interstellar medium (ISM) of the host galaxy. At higher radio frequencies, distinctive bright hot spots are found at the interface with the external ISM in the host galaxy \citep[e.g.,][]{Wil94,Tay96}. In some CSOs, the core is clearly visible because the emission from the core is Doppler boosted \cite{Pec00,Gug05}, whereas in most CSOs the central cores are too weak to be detected. A possible reason for the dimness of the core is the strong absorption by a dense circum-nuclear structure. Observations of broad H~I absorption lines (e.g., 1946+708: Peck et al. 1999; 0108+388, OQ208: Orienti, Morganti \& Dallacasa 2006b) and free-free absorption (e.g., OQ208: Kameno et al. 2000; 0108+388: Marr et al. 2001; 1511+0518: Orienti \& Dallacasa 2008) in some CSOs are in support of the existence of such an obscuring disk or torus. In a few other cases, the neutral and ionized gas in the innermost regions of the compact radio sources is in-homogenous and clumpy, as evidenced by interactions between the jet and off-nuclear clouds \cite{Ver03,Mor04,Mor05,Gup06,Lab06}. The integrated radio spectra of CSOs often show a peak around a few GHz frequency \cite{Mur03,Yang05}, which makes them part of the class of Gigahertz Peaked Spectrum (GPS) sources \citep[reviewed by][]{ODea98,Fanti09}. 
Many of these larger-size GPS galaxies are also CSOs with compact double or triple source structures.

The sub-galactic source structure of CSOs poses some most intriguing questions.  The symmetric double morphology and steep spectral index of CSOs are analogous to those of classical double radio sources \citep[e.g., FR II galaxies;][]{FR74}, but their sizes are 3--4 orders smaller than FR II galaxies. The "youth scenario" \citep[e.g.,][]{Phi82} considers that the small sizes of the CSOs result from their early evolutionary stage, and that they could eventually evolve into Mpc-scale extended radio sources. An alternative model, named the "frustration scenario", attributes the compactness of CSOs to retardation of the source expansion by the dense ISM within the host galaxy (e.g. O'Dea et al. 1991; Carvalho 1994, 1998). In the "frustration" scenario, GPS radio sources may never evolve into classical doubles. However, the physical properties such as the density structure of the dense ISM that confines these CSOs cannot be easily detected.

The two models predict distinctly different evolutionary fates and age determinations of CSOs provide a direct approach to discriminate between them. A ''kinematic age'' of CSOs may be derived from the propagation speed of the terminal hot spot with respect to the core and the projected distance of the hot spot from the core. In the absence of a visible core, the expansion speed of the hot spots is usually measured as the separation velocity between two hot spots using the brightest hot spot as the reference. Assuming that the two-sided jets expand with a same constant velocity, the kinematic age can be straightforwardly determined from multiple-epoch data. In addition, a ''radiative age'' of CSOs can also be estimated from their spectral break resulting from synchrotron aging of the relativistic electrons \cite{Mur99} assuming an equipartition magnetic field. In some CSOs with both kinematic and radiative age measurements, the two age estimates show good agreement ({\it e.g.}, 1943+546: Polatidis \& Conway 2003; Murgia 2003; 2352+495: Polatidis \& Conway 2003; Murgia et al. 1999).

The kinematic ages of CSOs derived from previous VLBI observations (Owsianik \& Conway 1998; Owsianik et al. 1998; Taylor et al.2000; Polatidis \& Conway 2003; Gugliucci et al. 2005; Polatidis 2009) range between 20 and 3000 years and provide strong evidence that some CSOs are really young radio sources. The histogram of CSO kinematic ages by Gugliucci et al. (2005) already shows an overabundance of ages lower than 500 year, suggesting part of the population consists of short-lived sources. Unfortunately only about two dozen CSOs with proper motion measurements are available to date and many of these are upper limits (Gugliucci et al. 2005; Polatidis \& Conway 2003; Giroletti \& Polatidis 2009). A large sample of CSOs with accurate proper motion measurements is critical for determining the age distribution, and for studying the dynamical evolution and physical properties of the CSOs in a statistical manner. 

In addition, Stanghellini et al. (2009) recently revealed non-radial motion of the hot spots in three CSOs, which has been interpreted as a result of precessing jets and provides an evidence of jet confinement by the external medium. If the CSOs do not have a long-term nuclear activity and/or the jet is not powerful enough to break through the confinement of the dense ISM, the growth of the radio source will be smothered and the CSOs appear young but are intrinsically much older. High-precision proper motion measurements are needed to test for the existence of the frustrated objects using the transverse motions of hot spots, since the non-radial motion will not be as distinct as the radial motion.

In order to establish a more complete sample of CSO proper motion detections, we carried out VLBI observations of a sample of ten CSOs and CSO candidates in 2005 and 2009. The observations in 2005 were made with the VLBA at five frequencies of 1.7, 2.3, 5.0, 8.4 and 15.4~GHz with the major purpose to identify the CSOs and to locate the cores by virtue of simultaneous radio spectra. The observations in 2009 of four of the ten CSO candidates at 8.4 GHz using a combined VLBI array consisting of four Chinese and two European telescopes were aimed at determining the proper motions of the hot spots and internal jet components in the CSOs.

The current paper is the first of a series of studies and focuses on proper motion measurements of the ten sources based on our observations in 2005 and 2009 and in combination with available VLBI archival data.
The earliest VLBI observations of the sample sources trace back to 1993, allowing us to better constrain the separation velocities of hot spots with a maximum time baseline of about 16 years. 
The results from the multiple-frequency VLBA observations in epoch 2005 will be presented in Yang et al. (in prep.; hereafter Paper II), and the detailed discussion of the physical properties and dynamical evolution of the CSOs will appear in An \& Baan (in prep., Paper III).  
In Section 2 we describe the sample selection criteria and the observational data from in 2009. The observing results and error analysis are presented in Section 3. In Section 4 we comment on individual sources. The main results are summarized in Section 5. Throughout this paper, we assume H$_0$ = 73 km s$^{-1}$ Mpc$^{-1}$, $\Omega_M$= 0.27, and $\Omega_\Lambda$= 0.73. The radio spectral index is defined as $S_\nu \propto \nu^{-\alpha}$.

\section{OBSERVATIONAL DATA}

\subsection{Sample Selection}

CSOs are found to be a significant occurrence within the GPS galaxy sample \citep{Xiang02,Ori06a}. The GPS sample is an ideal source to provide CSOs or CSO candidates since GPS sources are relatively easily identified in multiple-frequency spectra from single-dish or interferometer observations. Our CSO candidate sample consists of ten GPS sources with characteristic CSO morphology (symmetric compact double or triple sources) selected from a complete northern-hemisphere sample of GPS sources \cite{Mar99}. These ten sources were observed using the VLBA at five frequencies in 2005 with the major purpose of determining spectral indices of the VLBI components and a secondary purpose of measuring hot spot separation velocities in combination with previous VLBI observations. Four sources (J0132+5620, J0518+4730, J1734+0926, J2203+1007) are also included in the COINS \citep[CSOs Observed in the Northern Sky;][]{Pec00} sample and have been previously observed by the VLBA during 1997 and 2002 \cite{Pec00}. A brief description of the sample sources is given in Table \ref{tab:sample}.

Four high-declination CSO candidates (J0132+5620, J1324+4048, J1756+5748, J2312+3847) from our sample were further observed at 8.4~GHz in 2009 using a VLBI array consisting of the four stations of the Chinese VLBI Network (CVN) and two stations of the European VLBI Network (EVN). With the new observations, we aim to determine or to better constrain the angular separation velocities of hot spots in these CSOs and candidates. These four sources have been selected using the following selection criteria:

\noindent
(1)
All four sources are located in the northern sky with $\delta>+30\degr$, so that observations of high-declination sources with the Chinese and European antennas provide optimized ({\it u,v}) coverage for mapping the fine structure of the CSOs;
\\
(2)
All sources show simple double or triple structures on milli-arcsecond (mas) scales in previous epoch VLBI images;
\\
(3)
No proper motions and kinematic ages have been reported in these sources yet. 
The earliest VLBI observations of the CSO candidates in the current study were done in the early 1990s. Assuming a typical angular resolution of $\sim$1~mas of the observations and a signal-to-noise ratio of 20 for the hot spots, a proper motion accuracy of $\sim$1.3 $\mu$as yr$^{-1}$ may be achieved over a 16 year time span.
\\
(4)
All sources have a peak intensity higher than 100 mJy at 8.4 GHz. The high brightness of the sources guarantees a high position accuracy in an individual epoch. Because the sources themselves can serve as fringe fitting calibrators, most of the observing time may be spent on the target sources themselves.

\subsection{Observing Procedures in 2009}
\label{section:obs}

\begin{figure*}
\begin{center}
\includegraphics[scale=0.75,angle=-90]{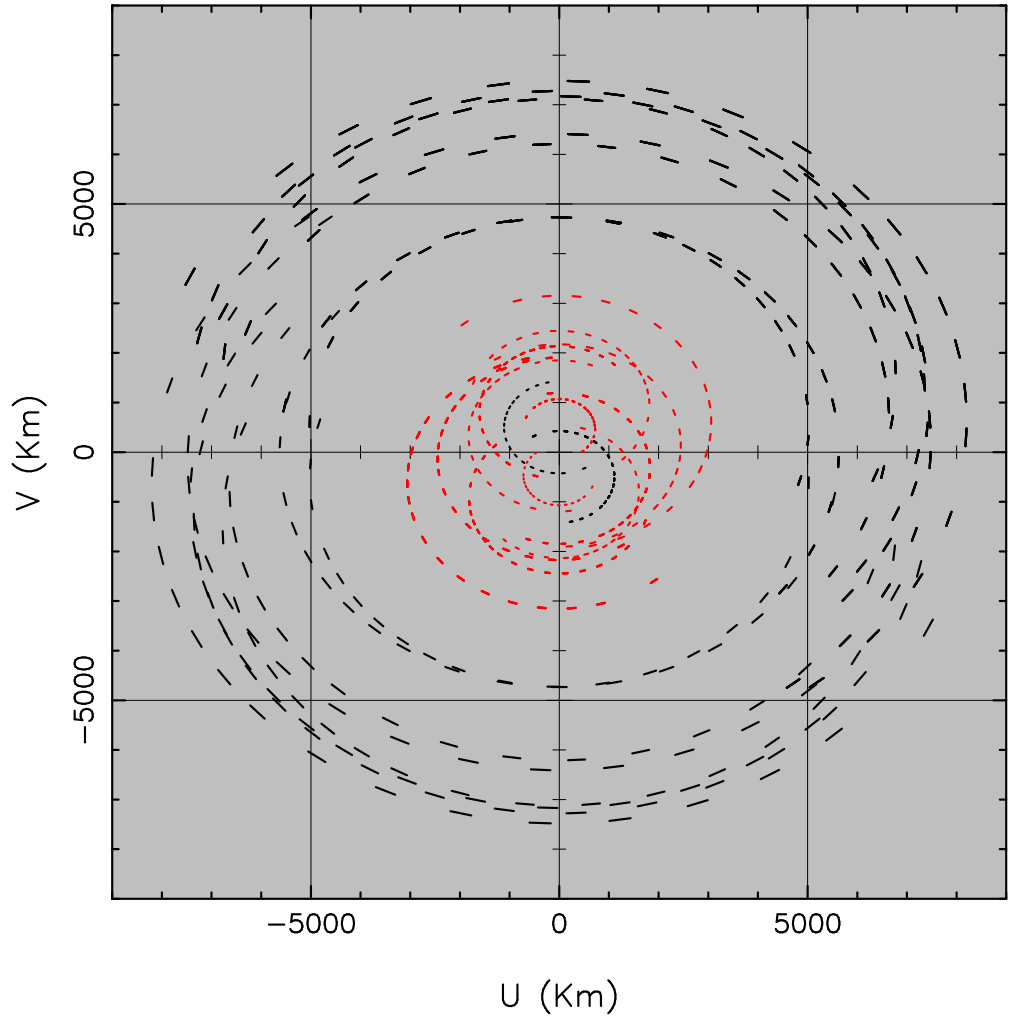}
\caption{ The ({\it u,v}) coverage of J0132+5620 derived from the present observations in 2009. The red dashed lines show the ({\it u,v}) coverage of only the Chinese telescopes. The black dashed lines show the baselines between Chinese and the two European telescopes. The inclusion of Miyun and Kunming telescopes into the EVN improves the long-baseline ({\it u,v}) coverage significantly. That is very crucial for mapping fine structures of compact sources.}
\label{fig:uvcov}
\end{center}
\end{figure*}

The observations of the four CSO candidates were carried out at 8.4 GHz on 2009 August 5 using four stations of the CVN, the Kunming 40m, Miyun 50m, Shanghai 25m, and Urumqi 25m radio telescopes, and two stations of the EVN, the Medicina 32m and Onsala 20m radio telescopes. The observing run lasted 24 hours. Since the CSO sources are strong enough at centimeter wavelengths to solve for fringe fitting solutions, we did not observe additional calibrators except for interspersing a few scans on fringe finders 4C~39.25 and DA~193. The effective on-source time was 3--4 hour for each of the target sources, which was split into about 18 scans. The observations of the four sources was interleaved and spread out in hour angle so as to acquire good ({\it u,v}) coverage. Figure \ref{fig:uvcov} shows an example of the ({\it u,v}) coverage of the observations. Right circular polarization (RCP) mode was used at all stations. The data were recorded in eight continuous channels of 8~MHz each, forming a total bandwidth of 64~MHz. Except for the Miyun and Kunming stations that recorded with 1 bit sampling, the other four stations used 2 bit sampling. The raw data were correlated at the Joint Institute for VLBI in Europe (JIVE), the Netherlands.

The post data reduction was carried out in the {\rm AIPS} software package \cite{Gre90} following the standard procedures (see the {\rm EVN} data analysis guide\footnote{http://www.evlbi.org/user\_guide/guide/userguide.html}). The amplitudes of the visibility data were calibrated using the system temperatures measured during the observations and the antenna gains. The visibility amplitudes of the calibrator DA~193 on each baseline were used to refine the amplitude scales for the Miyun and Kunming data. The parallactic angles were determined for each telescope and then used to correct the visibility phases. 4C~39.25 and DA 193 were applied to calibrate the complex bandpass response. Fringe fitting on DA~193 over a 10-min time span was used to align the phase delays between the channels. Global fringe fitting over the whole observing time with a solution interval of 2 minutes was used to solve for the delays and phase rates. After fringe fitting, the derived gain solutions were applied to the visibility data, which then were split into single-source data sets. The single-source data were imported to the DIFMAP software package \cite{She97,Pea94} to perform editing, self-calibration and imaging. For each source, a few iterations of phase-only self-calibration were applied to remove antenna-based residual phase errors. Next a few iterations of both amplitude and phase self-calibration were applied to eliminate amplitude calibration errors and to improve the dynamic range of the final image.

\section{RESULTS AND ANALYSIS}
\label{result}

\subsection{Images}

The analysis of the proper motions of the ten CSO candidates incorporates the data obtained from our observations in 2005 and 2009 as well as archival VLBA data, mainly from the VLBA Calibrator Search program \citep[{\rm VCS}:][]{Bea02,Pet08} and the {\rm COINS} program \cite{Pec00}. The total intensity images are displayed in Figures \ref{fig:j0017}--\ref{fig:j2312}. Table \ref{tab:map} lists the parameters of the images, including the restoring beam, the peak intensity ($I_{peak}$), the {\it rms} noise ($\sigma_{rms}$), and the contour levels. 

Two sources (J1734+0926 and J2203+1007) have two compact components with sharp boundaries at the outer edges and with nearly equal brightness at 8.4 GHz. The simultaneous multiple-frequency observations indicate a steep spectral index for the symmetric components (Paper II), which identifies them as CSO hot spots. J1324+4048 is also characterized by two compact steep-spectrum components and is identified as a CSO candidate.

Four sources (J0132+5620, J0518+4730, J1335+5844, and J1511+0518) display a triple structure with two bright terminal hot spots and a central component located approximately at the geometric center. Contrary to the other three triple-morphology sources, the central component in J0518+4730 is the brightest one. While the central component in J1511+0518 is identified as the flat-spectrum core, the identification of the central components in the other three awaits more observations with appropriately high resolution and sensitivity. More detailed discussion of the radio structures will be given in Section \ref{section:individual}.

\subsection{Model fitting and error analysis}

The quantitative study of the kinematics of the compact VLBI components employed Gaussian model fitting with the visibilities using the program {\rm MODELFIT} in {\rm DIFMAP}. Elliptical Gaussian components were fitted and subtracted in sequence until the residual peak intensity in the map was lower than 5 times the {\it rms} noise. Components that were too extended and complex were fitted with circular Gaussians taking more care of their peak positions. The fitting of components with a deconvolved size much smaller than the restoring beam always degenerated into a linear structures. These sources were fitted with circular Gaussians and an upper limit was set to the source size following the approach described in \citep{Lob05}. 
The identification of components is based on the (accurate) determination of their positions at all epochs. Multiple emission components having a similar size and flux density and a smaller difference in position between adjacent epochs have been identified as the same component.  

Table \ref{tab:modfit} lists the fitted parameters with the uncertainties given in brackets. The uncertainty of the integrated flux density ($S_{int}$) of the Gaussian component arises from two components, the measurement error ($\sigma_{M}$) and calibration error ($\sigma_{C}$). The measurement error $\sigma_{M}$ of the integrated flux density is described by the post-fitted {\it rms} fluctuations ($\sigma_{m}$) in the residual map multiplied by the apparent source size $\theta_{app}$ versus the restoring beam, {\it i.e.}, $\sigma_{M}= \frac{\theta_{app}}{\theta_{FWHM}}\sigma_{m}$. The error $\sigma_{M}$ is very sensitive to the apparent source size: for compact components it roughly equals the {\it rms} noise, while for weak and extended components it may become much larger. The calibration error $\sigma_{C}$ is determined from the calibration of the visibility amplitude. 

The amplitude calibration for the VLBI data was determined from system temperatures at two-minute intervals during the observations together with the antenna gain curves provided by each VLBI station. The amplitude calibration error may be estimated from the telescope amplitude correction factors during amplitude self-calibration. In most sources, this calibration error accounts for a few percent of the integrated flux density. Because of the diversity of the antenna performance of the Chinese and European antennas and the absence of large sensitive telescopes in the present observations, a conservative value was adopted for the average amplitude calibration uncertainty of 6 percent for the 8.4-GHz data. For the {\rm VLBA} data used in this analysis, the amplitude calibration uncertainty was assumed to be 5 percent at 5 GHz, 6 percent at 8.4 GHz and 7 percent at 15 GHz {\rm VLBA} calibration manual \cite{Ulv07}. The final error for the integrated flux density of the Gaussian component is expressed as $\sigma_{S_i} = \sqrt{\sigma_{M}^2 + \sigma_{C}^2}$.

The first order position uncertainty of the fitted components is represented by the ratio of the size of the restoring beam to the signal-to-noise ratio of the fitted component, $\frac{\theta_{FWHM}}{2 SNR}$, where $SNR=I_{peak} / \sigma_m$ \cite{Fom99}. Since the CSO core is always very weak, the relative proper motion of a jet or hot spot is actually measured with respect to a reference, which is often the brightest hot spot. Small deviations have been found for the fitted peak positions of the brightest VLBI components from the image centers at a level of a few tenths of a pixel size. This deviation of the peak position away from the image center has been treated as a systematic error of the reference point, which is included into the position errors of other components. For bright and compact components, the position error propagated from the reference point dominates over the position error defined by the signal-to-noise ratio. On the other hand, the first term of the position errors ($\frac{\theta_{FWHM}}{2 SNR}$) becomes dominant in extended components. The uncertainty of the separation $\sigma_{R}$ equals the position error, and the uncertainty of the position angle $\sigma_{\theta}$ has been defined as $atan(\frac{\sigma_R}{R}$).

\subsection{Proper motion determination}

The proper motions of VLBI components are determined as the rate of change in the separation of the component away from a reference point with time. The core is used as the reference when it is sufficiently bright and well confirmed. Otherwise, one of the terminal hot spots is used as the reference. For the sources with more than three-epoch data, linear regression fitting has been made to obtain the proper motion and also the epoch of zero separation. For sources with two-epoch data, the proper motions are calculated directly from the two data points. The proper motions are first determined separately along the relative ascension and declination directions ($\mu_\alpha$ and $\mu_\delta$), respectively. The motions of ($\mu_\alpha, \mu_\delta$) of each component are then converted to a representation along and perpendicular to the jet direction, {\it i.e.}, ($\mu_r, \mu_t$) using a Cartesian reference frame: $\mu_r = \mu_\alpha \sin(PA) + \mu_\delta\cos(PA)$, $\mu_t = -\mu_\alpha \cos(PA) + \mu_\delta\sin(PA)$. A mean position angle of the jet or hot spot is used in the calculations and the uncertainty in ($\mu_r, \mu_t$) is derived from the fitting errors of ($\mu_\alpha, \mu_\delta$) following standard error propagation. Tables \ref{tab:pm} and \ref{tab:pm2} present the proper motion measurements above 2$\sigma$.  

\subsection{CSO identification}

A CSO is strictly defined by the presence of a compact, flat- or inverted-spectrum core located between two steep-spectrum jets or outer hot spots of scale size less than 1~kpc \citep{Wil94,Pec00}. 
In this paper, the sources are identified as CSOs using  two major criteria: 
\newline (1) {\it Morphology and Spectral Index} -
The source is characterized by two symmetric, compact steep-spectrum lobes or hot spots, and in some cases, the central flat-spectrum core is also visible. The spectral indices of the VLBI components are derived from the multi-frequency data in epoch 2005 (Paper II); 
\newline (2) {\it Kinematics} A mirror symmetric motion pattern of the hot spots relative to a central position, even if there is no radio emission detected, is indicative of the existence of a central core in a CSOs. Otherwise, a flat-spectrum component appearing at one end of the radio structure (as in radio-loud blazars) is classified as a "core-jet" source. In addition, for CSOs with visible cores, flux variability is also used as a cross check for core identification. 

Following the above criteria, J1756+5748 and J2312+3847 (core-jet sources) have been ruled out for the CSO sample. The nature of J0017+5312 is considered doubtful. J1324+4048 is identified as a CSO candidate and the remaining six sources are identified as CSOs. The CSO sources J0132+5260 and J1511+0518 are confirmed as such for the first time. 

\section{DESCRIPTION OF INDIVIDUAL SOURCES}
\label{section:individual}

\subsection{J0017+5312 (Core-Jet)}
\label{section:j0017}

\begin{figure*}
\begin{center}
\includegraphics[scale=0.65]{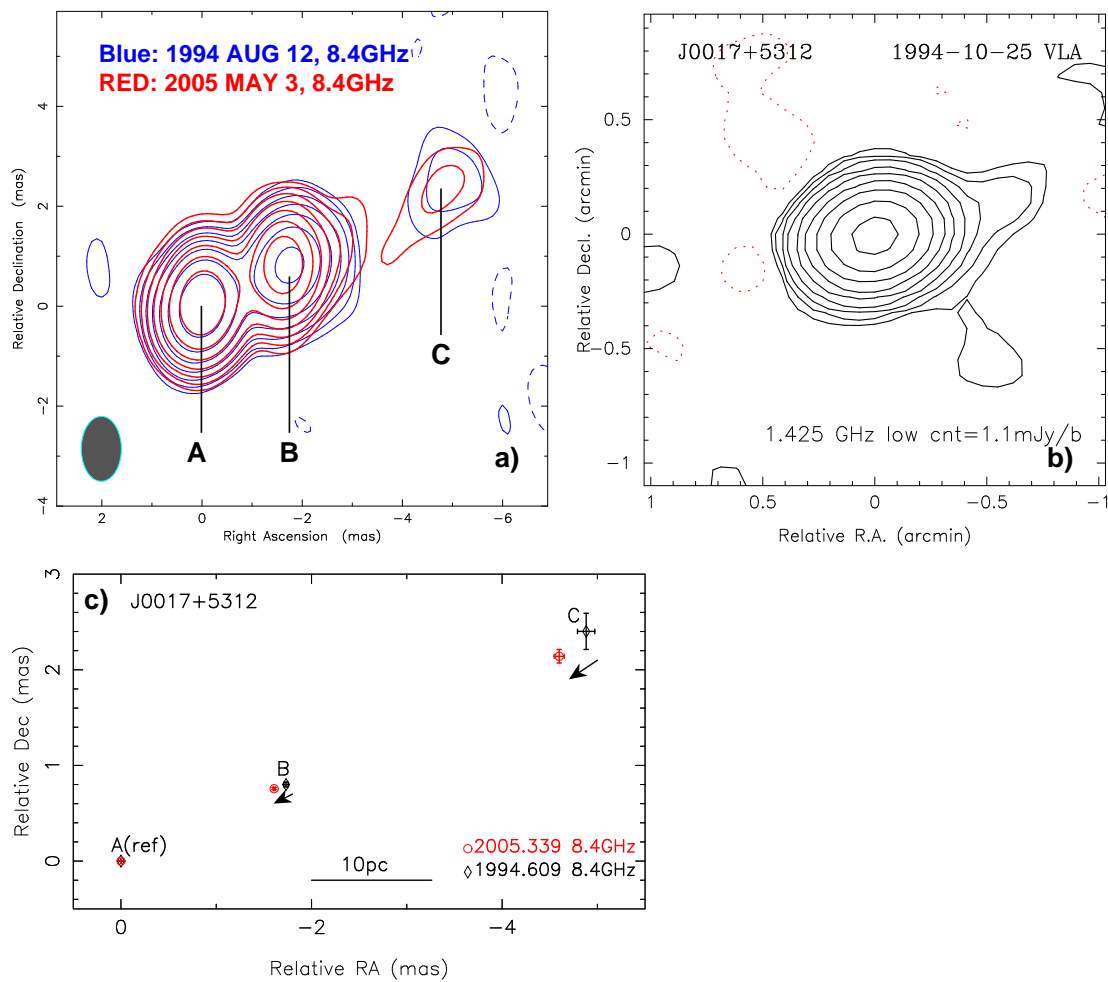}
\caption{The structure of Core-Jet source J0017+5312 with the core hidden at the eastern end of component {\rm A}.  {\it a}) Overlay of total 8.4 GHz intensity images of J0017+5312 on 1994 August 12 (thin contours) and 2005 May 3 (thick contours). {\it b}) The 1.5-GHz VLA image of J0017+5312 using a restoring beam of $20.2\arcsec\times14.9\arcsec$ (PA=$-75\degr$). The contours are 1.1$\times$($-$1,1,2,4,...256) mJy beam$^{-1}$. {\it c}) The 2-D positions of the VLBI components. {\rm B} and {\rm C} show a motion towards the {\rm A}, the reference point.}
\label{fig:j0017}
\end{center}
\end{figure*}

The core-jet sourceJ0017+5312 is associated with a quasar at redshift $z=2.574$ (1 mas = 7.926 pc) \cite{SE05}. Its total flux density spectrum turns over at around 5 GHz \cite{Mar99}. The 1.5-GHz NVSS image shows a compact component with a flux density of $\sim$380 mJy and a weak extension $\sim$30\arcsec{} ($\sim$240 kpc) northwest of the core (Figure \ref{fig:j0017}-b) with a peak intensity of about 8 times the {\it rms} noise. However, this extended component is not detected in the VLA images at 8 and 43 GHz (VLA/NRAO archive) suggesting a steep spectrum. This extended feature may represent the relic from nuclear activity at a previous epoch. The morphology of a compact jet along with a distant extended lobe is common in core-jet sources in which the advancing jet is Doppler boosted, and it is also observed in other CSOs (J0108+388; Baum et al. 1990). In the 2.3-GHz VLBA images, J0017+5312 shows a slightly resolved structure with a projected source size of $\sim$40 pc (VCS archive).

As an example, Figure \ref{fig:j0017}-a shows the source structure derived from the 8.4-GHz observations in 1994 and 2005. Two compact components {\rm A} and {\rm B} dominate the total flux density, while the extension {\rm C} to the northwest (PA$=-65\degr$) aligns with the VLA extended feature. 
Both {\rm A} and {\rm B} show inverted spectra with a turnover between 5 and 8 GHz, and the high-frequency spectral index, $\alpha^{15.4}_{8.4}\approx 0.7$, identifies both as steep-spectrum components. The weak component {\rm C} has a much steeper spectrum with $\alpha^{15.4}_{1.7}=1.1$ (Paper II).
Although the 8.4-GHz radio structure of J0017+5312 has the morphology of a typical CSO, the absence of a flat-spectrum central core renders the CSO identification less certain. 
The spectral index map shows a hint of a flatter-spectrum region at the outer edge of the eastern lobe (Paper II), if it is not an artifact, invoking a core-jet interpretation of the source structure. In this scenario, the visible emission structure in VLBI images and the extended VLA feature are associated with a single-sided jet.
Similar to many other CSS quasars, the active nucleus of J0017+5312 is likely obscured by the bright innermost jet and the core itself is dimmed due to either synchrotron self-absorption or free-free absorption (as for the weak nucleus and the bright inner hot spot {\rm B} in 3C~48: An et al. 2010).
Although the possibility of a CSO classification for J0017+5312 is not fully excluded, we tentatively identify J0017+5312 as a core-jet source.

Besides the observational data in 2005, also the archival 8-GHz VLBA data of J0017+5312 observed in 1994 has been mapped. A direct comparison of the radio images in epochs 1994 and 2005 shows a shrinking of the source size (Figure \ref{fig:j0017}-a). The separation of {\rm B}--{\rm A} decreases by 0.132 mas and {\rm C}--{\rm A} decreases by 0.367 mas within a time span of 10.73 year. That gives rise to a relative proper motion of 12.3 $\mu$as yr$^{-1}$ ({\rm B}) and 34.1 $\mu$as yr$^{-1}$ ({\rm C}) toward {\rm A}, corresponding to apparent transverse velocities of $1.1\,c$ (B) and $3.2\,c$ (C), respectively. However, in analogy with previous reports of contraction of radio sources (e.g.,  J0650+6001: Akujor et al. 1996, Orienti \& Dallacasa 2010; J11584+2450: Tremblay et al. 2008), the apparent contraction in J0017+5312 may result from a bias resulting from the motion of the innermost jet, which is used as the reference for proper motion measurements. In this picture, the reference component {\rm A} in J0017+5312 is actually a combination of a bright moving component (jet knot) and a weak stationary component (core). A new and younger jet knot is created at a later epoch and an older hot spot has advanced out of the core and fades at a larger distance. Therefore, the apparent shrinking of the source size may actually reflect the motion of the internal jet knot and not the real contraction of the radio source. 

\subsection{J0132+5620 (CSO)}

\begin{figure*}
\begin{center}
\includegraphics[scale=0.65]{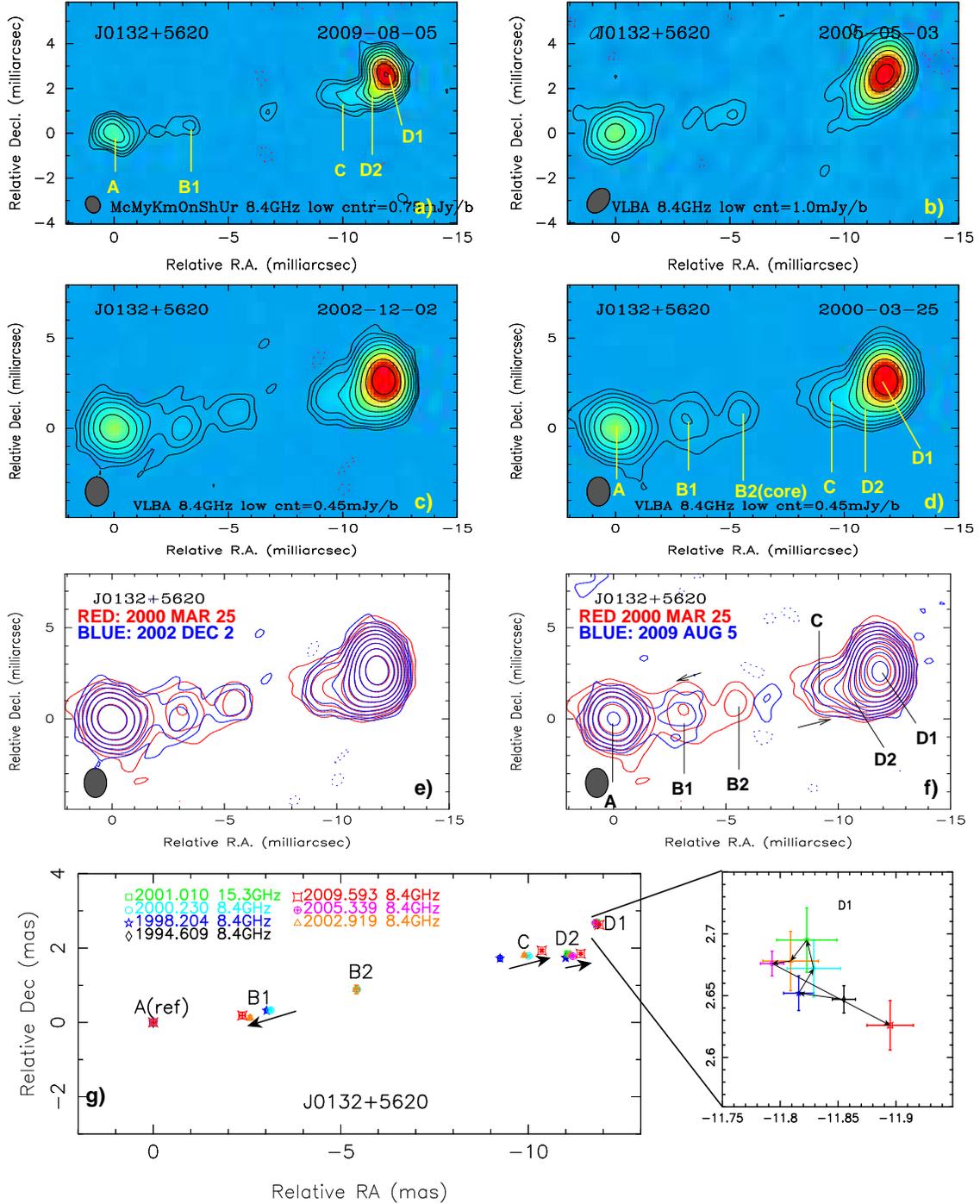}
\caption{The CSO source J0132+5620 with core component {\rm B2}. The source structure is presented at 8.4GHz at four epochs: {\it a)} The  epoch 2009 August 5 is from the current observations, {\it b)} The epoch 2005 May 3 taken from Paper II; {\it c)} The epochs 2002 December 2 and 2000 March 25 are derived from the VLBA archives. The overlay of images in different epochs (Fig.\ref{fig:j0132}-e and -f) presents changes of the radio structure. The image parameters are presented in Table \ref{tab:map}. Fig.\ref{fig:j0132}-g displays the 2-D distribution and proper motions of the VLBI components.}
\label{fig:j0132}
\end{center}
\end{figure*}

\begin{figure*}
\begin{center}
\includegraphics[scale=0.6]{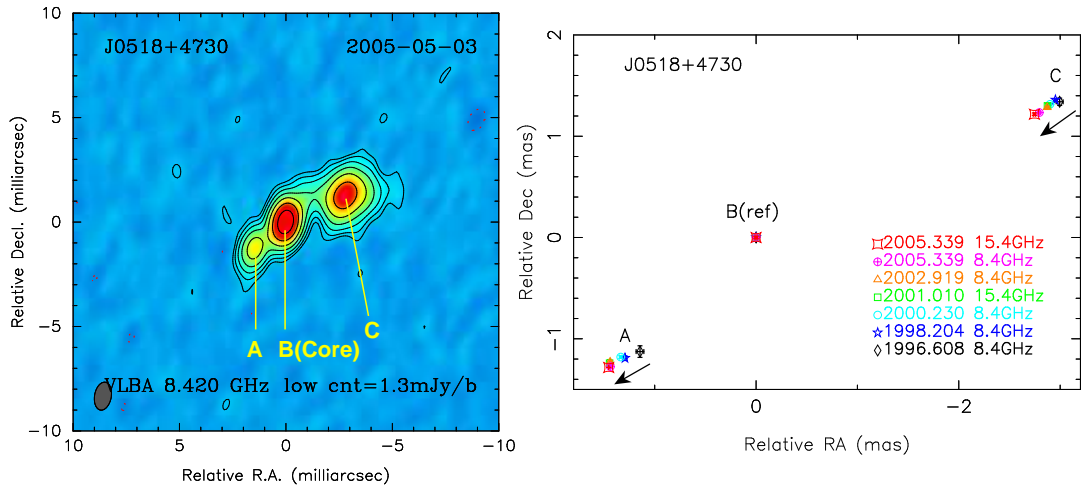}
\caption{The CSO source J0518+4730 with {\rm B} as core. {\it Left:} The 8.4-GHz CLEANed image of J0518+4730 at epoch 2005 May 3. {\it Right}: The 2-D distribution of VLBI components.}
\label{fig:j0518}
\end{center}
\end{figure*}

The optical redshift of the host galaxy of J0132+5620 is not available in the literature. This source was included in the {\rm COINS} sample and has been observed with $\sim$1\,mas resolution during 1997 and 2002 \cite{Pec00,Gug05}. It displays a symmetric double-source structure in the East-West direction at 2.3 GHz (VCS archive). 
Figure \ref{fig:j0132} shows the 8.4-GHz images of J0132+5620 derived from the new 2009 observations and the previous ones made in 2000, 2002 and 2005  (Peck \& Taylor 2000; Gugliucci et al. 2005; the current paper). The source components identified with model fitting (see Table \ref{tab:modfit}) are labeled in the images following the nomenclature used in Peck \& Taylor (2000). The emission structure is dominated by two hot spots {\rm D1} at the western end and {\rm A} at the eastern end of the radio source, respectively. The continuous emission structure connecting {\rm D1} and {\rm A} present in the 2.3- and 5-GHz images has been resolved at 8.4 GHz. Four internal jet components are detected in the intermediate region between {\rm A} and {\rm D1}. There is a weak component {\rm B2} midway between the two hot spots.  

The two lobes are separated by about 12\,mas, and the western one appears brighter with an intensity ratio $R_{W:E}\sim$1.2:1 at 2.3 GHz. Both lobes show a rising spectrum at lower frequencies 
(eastern lobe: $\alpha^{2.3}_{1.7}=-0.4$; western lobe: $\alpha^{2.3}_{1.7}=-0.4$) and a steep spectrum at higher frequencies (eastern lobe: $\alpha^{15.4}_{5.0}=3.0$; western lobe: $\alpha^{15.4}_{5.0}=2.2$) 
with turnovers around 2.3 GHz (the western lobe) and 5 GHz (the eastern lobe) (Paper II). At 5 GHz and higher frequencies, the western and eastern lobes are resolved into several sub-components. The simultaneous 5- and 8-GHz data in 2005 give rise to an upper limit of the spectral index of {\rm B2} $\alpha^{8.4GHz}_{5.0GHz}\lesssim0.0$ (Paper II). All other VLBI components show steep spectra. Since {\rm B2} is the only flat-spectrum VLBI component and lies near the geometric center of J0132+5620, it is most likely that {\rm B2} is associated with the core. Future simultaneous VLBI observations at 5 and 8.4 GHz with high sensitivity are needed to confirm the core identification of {\rm B2}.  

The flux densities of lobe-dominated CSOs usually do not vary rapidly. Except for the earliest (1994 August 12, or 1994.609) and the latest epochs (2009 August 5, or 2009.593), the flux densities of the two hot spots from epoch 1998.204 to epoch 2005.339 show a variation of less than 5 percent (Table \ref{tab:modfit}). This variability level is within the amplitude calibration uncertainty of the VLBA data. The increase of the flux density of {\rm D2} and {\rm A} in epochs 2009.593 and 1994.609 may partly result from overestimated scaling of the visibility amplitude and/or mis-calibrated correction factors when using the amplitude self-calibration. The suspicious core {\rm B2}, due to its weakness, does not show any variability from the only two-epoch observations.

In combination with our observations of J0132+5620 in 2005 and 2009, six epochs VLBI data at 8.4 and 15.4 GHz have been collected from the NRAO archive with a maximum time span of $\sim$15 years. Figures \ref{fig:j0132}-e and \ref{fig:j0132}-f superpose the total intensity images derived from the observations in 2002 and 2000, as well as in 2009 and 2000, respectively. The images have been re-created with the same cutoffs of the ({\it u,v}) range, the same cell size and the same restoring beam. The source structures revealed by the three data sets are exactly consistent with each other, except that the 2009 image does not detect the central component {\rm B2}. The direct comparison of the images in different epochs suggests that the separation between the two hot spots, or the overall source size, does not change significantly over the 15-year time span. In contrast, the internal jet component {\rm B1} shows a clear motion to the East and the centroid of {\rm C} shifts to the West.

The available data have similar ({\it u,v}) coverage and angular resolutions and reveal self-consistent source structures, which allows for a linear regression analysis of the proper motions of VLBI components. As mentioned above, the suspicious core {\rm B2} is so weak that its position uncertainty accounts for (15--30)\% of the beam size. Such a large position uncertainty prohibits using {\rm B2} as the reference point in proper motion measurements. Instead the compact and bright hot spot {\rm A} has been used as a reference to determine the expansion rates of other components. Figure \ref{fig:j0132}-$g$ displays the 2-dimensional distribution of the VLBI components. The source exhibits a nearly aligned structure along a mean P.A. = $-80.7\degr$. Core component {\rm B2} does not show significant positional changes with respect to {\rm A} between the epochs 2000 and 2002. VLBI monitoring over an even longer time span is necessary to detect any small change of the separation between {\rm B2} and {\rm A}. The hot spot {\rm D1} at the western extreme of the source is not detected with any clear expansion along the jet axis either. On the contrary, {\rm D1} shows a motion to the northeast from epoch 1994 to 2005, in a direction perpendicular to the jet axis. Without including the data point in epoch 2009, which shows large deviation away from the other positions on earlier epochs, the linear regression fitting gives rise to the proper motions of {\rm D1}: $\mu_r=-4.9\pm1.3$ $\mu$as yr$^{-1}$ along the jet axis (toward the reference point), and $\mu_t=-3.7\pm1.3$ $\mu$as yr$^{-1}$ perpendicular to the jet direction. Using the 2-$\sigma$ proper motion uncertainty ($\sim$2.6 $\mu$as yr$^{-1}$) as an upper limit of the hot spot expansion rate, a lower limit of $4700/(1+z)$ year is found for the kinematic age of J0132+5620.

The other three jet components show proper motions in good alignment with the jet axis. This is in agreement with the general picture of young FR~II sources in which the jet material ejected from the central engine feeds the lobes through a channel in which the interstellar medium is swept by the expansion of the hot spots/lobes. Because the hot spot {\rm A} and the core {\rm B2} does not show significant relative motion, the derived proper motions of jet knots can be approximately converted to the velocities relative to the core so as to investigate the variation of jet velocities with the increasing radial distances. The magnitude of the proper motion of the jet components shows a decreasing trend with increasing separations from the core, i.e., $\mu_r=72.7\pm6.8$ $\mu$as yr$^{-1}$ ({\rm B1-B2}, $d\sim3.0$ mas), $\mu_r=59.6\pm4.0$ $\mu$as yr$^{-1}$ ({\rm C-B2}, $d\sim5.0$ mas), and $\mu_r=27.8\pm1.8$ $\mu$as yr$^{-1}$ ({\rm D2-B2}, $d\sim6.0$ mas). Such a systematic change in the jet proper motions does not likely result from geometric projection effects, since this source lies close to the plane of the sky. Moreover it seems unrealistic that the initial speeds of the jet knots change significantly within a few years. 

A possible interpretation of this systematic change is that the jet components are ejected from the central nucleus at a constant 
initial velocity, and they experience a deceleration process before reaching the terminal hot spot. A simple picture of the motion of a jet knot or condensation in the underlying jet flow calls for a balance of its internal jet pressure with the ram pressure $\rho_e v_j^2$, where $\rho_e$ is the density of the ISM surrounding the jet, and $v_j$ is the advancing speed or equivalently the bulk velocity of the jet knot. Assuming that the jet thrust ($F_j=\pi r_j^2 \rho_e v^2_j$) remains constant and the density $\rho_e$ is a weak function of the distance from the core, then the variation of the advancing speed of the jet knot is correlated with the change of the radius of the jet knot. It is reasonable to expect an increasing trend for the component size of the jet knot during its propagation outward, as a result, the advancing speed of the jet knot decreases with the increasing distance accordingly. Alternatively, the internal jet pressure remains unchanged and the jet is confined by ram pressure. The density of the ISM surrounding the jet knot will then increase with distance as the jet approaches the lobe due to enhanced back-flow. As the result, the jet experiences a deceleration when passing through the denser ISM.

The mirror symmetry of both the radio morphology and jet motions provides strong evidence for the presence of a central core between {\rm B1} and {\rm C}, most likely associated with {\rm B2}. The present kinematic study confirms the CSO identification of J0132+5620.

Assuming that the brighter hot spot {\rm D1} is associated with the advancing lobe and {\rm A} 
represents the terminal hot spot in the receding lobe, we may derive constraints on the kinematic 
parameters of the jet from the intensity ratio $R_i$ of the advancing and receding jets:
\begin{equation}
R_i=\frac{S_{adv}}{S_{rec}} = (\frac{1+\beta\cos\theta_v}{1-\beta\cos\theta_v})^{3+\alpha}
\label{equ:iratio},
\end{equation} 
where $S_{adv}$ and $S_{rec}$ represent the flux density of the advancing and receding hot spots,  $\theta_v$ is the viewing angle between the jet axis and the line of sight, and $\alpha$ is the spectral index of the jet, defined as $S\propto\nu^{-\alpha}$. Assuming the advancing and receding jets are ejected at an equal speed and the intensity difference between 
hot spots is solely attributed to the Doppler boosting effect, the calculations using the intensity 
ratio of $S_{D1}/S_{A}$ give rise to  $\beta\cos\theta_v\sim0.24\,c$. This number suggests 
mildly relativistic jet flow moving at a modest viewing angle.
The redshift of J0132+5620 is not known yet, therefore the conversion from the angular size to 
physical size remains uncertain. Assuming a moderate viewing angle $\theta_v=45\degr$, we 
get a jet velocity $\beta=0.34\,c$ and an apparent velocity $\beta_{app}=0.32\,c$. The mean 
angular velocity of the internal jet knots is $53$ $\mu$as yr$^{-1}$, resulting in a conversion 
factor of 1~mas yr$^{-1}=5.9\,c$. This would place J0132+5620 in the nearby Universe at a 
redshift of $z\sim0.1$. Further optical spectroscopic observations of this source are necessary to 
verify the redshift of this source.

\subsection{J0518+4730 (CSO)}

CSO source J0518+4730 is another source from the COINS sample \cite{Pec00} but it has no known optical redshift. The VLBI image derived from the present 8.4-GHz data displays a triple structure (Figure \ref{fig:j0518}), in good agreement with previous observations at similar frequencies (Peck \& Taylor 2000). The eastern and western components show a slight misalignment with a difference of $\sim17\degr$ in position angle. Gugliucci et al. (2005) classified J0518+4730 as a core-jet source and identified the easternmost component {\rm A} as the core. 
However, the observations in 2005 indicate a steep spectrum for {\rm A} ($\alpha^{15.4}_{1.7}=1.2$) and {\rm C} ($\alpha^{15.4}_{1.7}=0.8$). The central component {\rm B} is the most compact component, but it also has a steep spectrum with $\alpha^{15.4}_{5.0}=0.7$ (Paper II). B is most likely associated with the innermost jet in the vicinity of the core. Therefore, J0518+4730 may be classified as a CSO with a triple morphology.

Using component {\rm B} as a reference, the proper motion of the hot spot {\rm A} is along the jet axis with an expansion velocity $\mu_r=29.8\pm2.9$ $\mu$as yr$^{-1}$. Component {\rm C} appears to move toward the core from epochs 1996 to 2005 at a rate of $25.7\pm1.9$ $\mu$as yr$^{-1}$. As discussed in Section \ref{section:j0017}, the apparent inward motion of the hot spot {\rm C} may be interpreted as a change in the emission structure of the hot spot within the lobe or the motion of the reference point {\rm B} itself. Both the flux density and component size of {\rm B} show a monotone increase at 8.4 GHz from epoch 1996 to epoch 2005. This provides a support for the creation of a new component at the jet base. When the newly formed jet component dominates the emission of the core, the propagation of the jet downstream would probably result in the observed shortening of the western arm {\rm B--C} if the reference point is bound to the moving component {\rm B}. If this scenario is correct, the observations suggest that the newly ejected jet knot {\rm B} moves at $\sim$27.8 $\mu$as yr$^{-1}$ to the southeast, and the two hot spots move at $\sim$2 $\mu$as yr$^{-1}$ away from the core assuming they have same expansion velocities. Inspection of the change of the {\rm A--C} arm length with time leads to an expansion rate of 3.6 $\mu$as yr$^{-1}$, in agreement with the above scenario that {\rm B} is a moving component in the vicinity of the core. The expansion velocity corresponds to a kinematic age of $\sim1200/(1+z)$ year. VLBI monitoring of the time evolution of the spectral index, the radio flux density, and the component size of {\rm B} will be needed to confirm the above interpretation of the jet kinematics and to further constrain the kinematic age of this source.

\subsection{J1324+4048 (CSO Candidate)}
\begin{figure*}
\begin{center}
\includegraphics[scale=0.75]{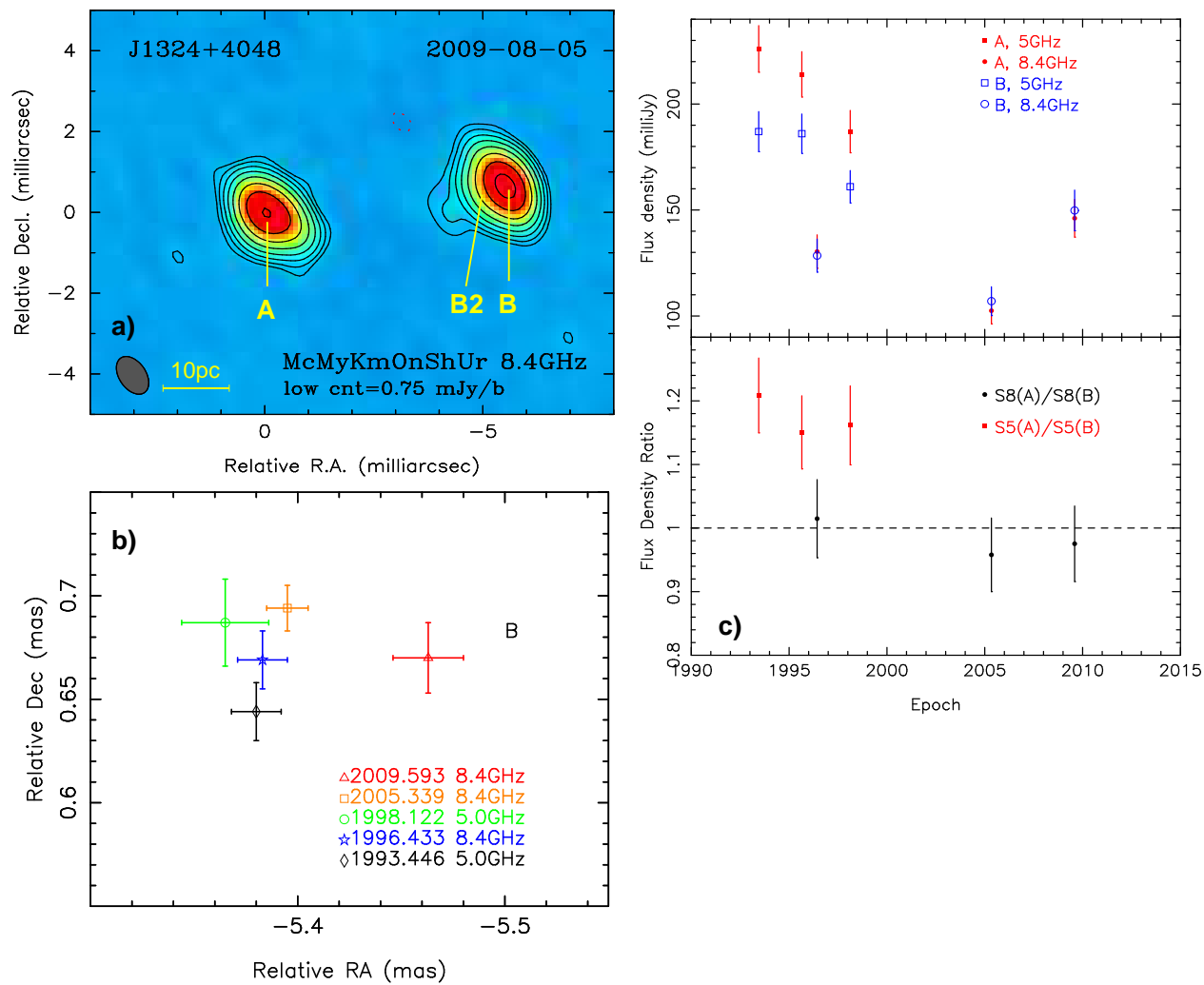}
\caption{The CSO candidate source J1324+4048 without an identifiable core component. {\it a)}: The 8.4-GHz CLEANed image of J1324+4048 at epoch 2009 August 5; {\it b)}: 2-D distribution of VLBI components; 
{\it c)}: The flux density variation of J1324+4048. The {\it Top} panel of Fig.\ref{fig:j1324}-$c$ gives the integrated flux densities of components A and B at different epochs, and the {\it Bottom} panel displays the variation of the flux density ratio of A and B. The flux density ratio shows a steady decrease with time}
\label{fig:j1324}
\end{center}
\end{figure*}

CSO candidate source J1324+4048 is associated with a quasar at redshift of 0.496(1 mas = 5.898 pc) \cite{Ver96}. Figure \ref{fig:j1324}-a displays the 8.4-GHz image observed on 2009 August 5. The source is characterized by two compact components along PA=$-83\degr$ with a separation of 32.4~pc, in agreement with previous VLBI observations at close frequencies \citep{Hen95,Bri08}. These two components show a steep spectrum, {\it i.e.}, $\alpha^{15.4}_{2.3}=1.2$ (eastern) and $\alpha^{15.4}_{2.3}=0.7$ (western) (Paper II). 
The symmetric morphology, edge-brightening and steep spectra of two components identify J1324+4048 as a CSO, even though a flat-spectrum core is not identifiable from these observations. However an alternative interpretation of a 'core-jet' source cannot be fully ruled out. In this scenario, the observed two components are bright knots in a single-sided jet, while the core, which is not detected in the present observations, lies at one end of the jet. In the current paper, J1324+4048 is regarded as a CSO candidate and the bright components are hot spots. 

Figure \ref{fig:j1324}-b shows the positions of {\rm B} with {\rm A} as the reference. The separation between the two hot spots {\rm A} and {\rm B} does not show a significant change between epochs 1993 and 2005. This is consistent with the nondetection of radial expansion between 1993 and 1998 \citep{Bri08}. {\rm B} appears to move in a loop-like trajectory: after moving to the north from 1993 to 2005 it abruptly turns to the southwest in 2009. This wandering of the hot spot is indicative that the jet head impacts on a wall of external medium at different sites before finding a way to breaks through the obstacle, in analogy with the picture of the 'dentist's drill' model, first proposed to explain the multiple hot spots in FR II galaxies \cite{Sch82}. In order to avoid systematic errors of component positions due to frequency-dependence opacity, only 8.4-GHz data were used for the proper motion analysis. Linear regression fitting to the position variations of {\rm B} with the time gives a radial expansion velocity of $v_{r}=0.12\pm0.04\,c$, resulting in a kinematic age of $870\pm290$ years

Another interesting result is that both {\rm A} and {\rm B} show variability on time scales of a few years (shown in Figure \ref{fig:j1324}-$c$). The flux densities of {\rm A} and {\rm B} at both 5 and 8.4 GHz decrease since epoch 1993 until they reach a minimum in epoch 2005, reflecting the fading of the hot spots due to adiabatic expansion. After that, the flux densities of both components shows a significant increase (by 40 percent) in epoch 2009. A jet component {\rm B2} is detected at the starting of the western mini-lobe since 1998. Comparison of the epochs 2005 and 2009 data suggests a superluminal velocity of $v_{r}(B2)=2.2\pm0.5\,c$ along the jet axis. The feeding of the relativistic jet flow into the lobe may be responsible for the flux density increase of the hot spot {\rm B}. In addition, the flux ratio of two hot spots ($S_A/S_B$) flips after epoch 1998. Before 1998, the ratio was $S_A/S_B=1.2$ in epoch 1993 and changed to $\sim$0.96 in epochs 2005 and 2009. The flux density ratio flip invokes different jet power of the two-sided jets or a different conversion efficiency from jet kinetic energy to radiation energy in two lobes.

\subsection{J1335+5844 (CSO)}

\begin{figure*}
\begin{center}
\includegraphics[scale=0.75]{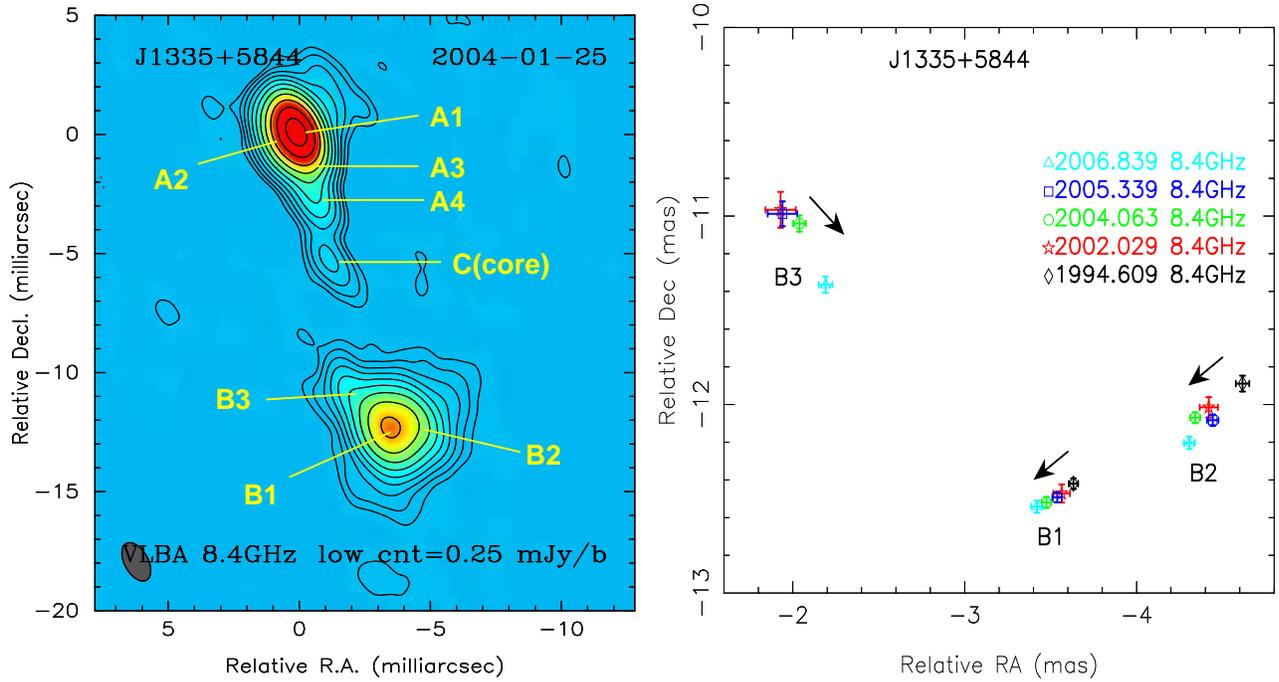}
\caption{The CSO source J1335+5844 with core component {\rm C}. {\it Left}: The 8.4-GHz CLEANed images of J1335+5844 at epoch 2004 January 25. {\it Right}: The 2-D distribution of VLBI components.}
\label{fig:j1335}
\end{center}
\end{figure*}

The host galaxy of J1335+5844 (also named as 4C $+$58.26, $m_V=22$) was found to be too faint to measure the redshift. 
Stanghellini et al. (2009) estimated a photometric redshift of 0.57 for this source (1 mas = 6.331 pc).
The source has been identified as a High Frequency Peaker (HFP) with a turnover frequency at about 5 GHz \cite{Dal00}. 
The 8.4-GHz image derived from the VLBA observation in 2004 (see also Fig. 1 in Dallacasa et al. 2005, and Fig. 2 in Stanghellini et al. 2009) is shown in the left panel of Figure \ref{fig:j1335}.  The VLBI image exhibits a triple source structure in a north-south direction with a total extent of $\sim$13 mas (84.5 pc). The northern component is characterized by a compact bright head which is fitted with three Guassian components  {\rm A1}, {\rm A2} and {\rm A3}. The northern component shows an elongated extension tracing back to the central component {\rm C}, which is clearly detected in high-dynamic-range images in epochs 2004 and 2006, whereas there is only a hint of emission ($\lesssim 3\sigma$) at other epochs. Comparison of the epoch 2005 8.4-GHz image and the tapered 15.4-GHz image gives a rough estimate of the spectral index of {\rm C} with $\alpha^{15.4GHz}_{8.4GHz}\lesssim 0.4$.  The recent multi-frequency VLBA observations of J1335+5844 made in 2010 confirm the flat spectral index of {\rm C} (D.~Dallacasa, private communication). Therefore, we consider J1335+5844 as a CSO in light of the compact symmetric structures with respect to a central component. 

The southern component appears more extended to the west and the northeast, where {\rm B2} and {\rm B3} are fitted, respectively. The brightest northern subcomponent {\rm A1} shows a relatively flat spectrum with $\alpha_{8.4GHz}^{15GHz}\sim 0.02$, in agreement with previous measurements \cite{Ori06a}. The 15-GHz data present a deconvolved source size of $0.29\times0.13$ (mas) for {\rm A1}, corresponding to a brightness temperature of $7\times10^{10}$ K. A fit using the synchrotron self-absorption model to the spectrum of the entire northern component ('A1+A2+A3') gives a turnover frequency as high as 11 GHz (Paper II). The high turnover frequency and high brightness temperature of the northern bright head {\rm A1} are consistent with an interpretation that {\rm A1} represents a very compact hot spot  \citep{Ori06a} and the flatness of the spectrum results from the injection of fresh particles from the central engine. If the spectrum turnover is dominantly due to synchrotron self-absorption, the observations would suggest a magnetic field with a strength of $B=0.05$ Gauss under the assumption of the equipartition between the magnetic field and particle energy. Such a high level of magnetic field strength seems to be common for HFPs \cite{OD08}.  

J1335+5844 was observed at 8.4 GHz at five epochs between 1994 August and 2006 November and may be used for proper motion analysis. We note that the model fitting of weak components (e.g., {\rm A2} and {\rm A3}) is susceptible to the internal emission structure change of the close-by brightest component. Their fitted parameters show large fluctuations and prevent any reliable proper motion calculation. Moreover, model fitting in the visibility domain is sensitive to the {\it uv} sampling. For example, the component {\rm A4} failed to obtain good fits in epochs 1994 and 2005 due to relative poor {\it uv} coverage. For these reasons, the proper motions are only computed for the bright and compact components {\rm B1}, {\rm B2} and {\rm B3}. The right panel in Figure \ref{fig:j1335} displays the positions of VLBI components {\rm B1}, {\rm B2} and {\rm B3} with the northern hot spot {\rm A1} as a reference. The inner jet {\rm B3} moves much faster than the terminal hot spots. The secondary hot spot {\rm B2} moves faster than the primary hot spot {\rm B1}, which only shows marginal expansion along the jet axis. Interestingly, two hot spots {\rm B1} and {\rm B2} exhibit a dominant motion in the direction perpendicular to the jet axis.  The present measurements of proper motions over a time span of 12.2 year confirm the transverse motion of the southern hot spots previously detected by Stanghellini et al. (2009) based on epochs 2004 and 2006 data.  The kinematic properties of the hot spots, {\it i.e.} the dominant transverse motion and non-detection of significant radial expansion, are also measured using the 5 GHz data over a time span of 4.2 year \citep{Bri08}. The deceleration of the terminal hot spot and the significant transverse motion likely manifest a confinement of the hot spot expansion by the interstellar medium. The radial velocity of {\rm B1}, $\mu_r = 4.7\pm3.0$ $\mu$as yr$^{-1}$ ($v_r=0.16\pm0.10\,c$), results in a kinetic age of $1800\pm 1150$ year. 

\subsection{J1511+0518 (CSO)}

\begin{figure*}
\begin{center}
\includegraphics[scale=0.75]{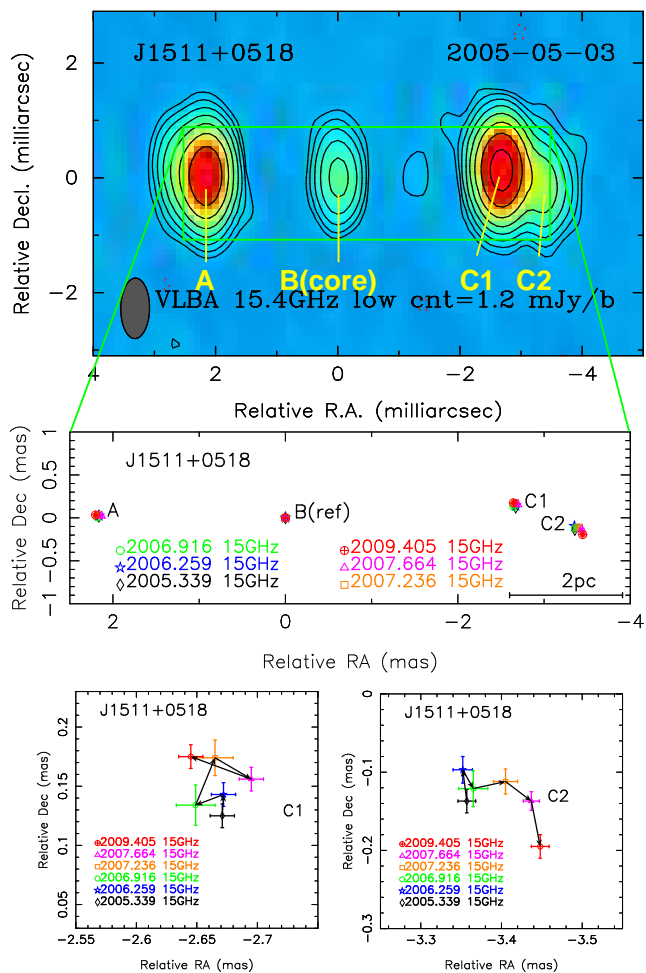}
\caption{The CSO source J1511+0518 with core component {\rm B}. {\it Top}: The 15.4-GHz CLEANed image of J1511+0518 at epoch 2005 May 3.
{\it Middle}: The 2-D distribution of VLBI components.
{\it Bottom}: The position variations of {\rm C1} and {\rm C2}. 
{\rm C1} shows sideways motions of the jet head, and {\rm C2} shows a motion toward the southwest as it appears to move away from {\rm C1}.
}
\label{fig:j1511}
\end{center}
\end{figure*}

CSO source J1511+0518 is associated with a Seyfert I galaxy at a redshift of 0.084 (1 mas = 1.527 pc; Chavushyan et al. 2001). It displays typical CSO morphological and spectral characteristics: the source is resolved into two hot spots {\rm A} and {\rm C} separated by $\sim$4.8 mas (or 7.3 pc) in East-West direction at 2.3 and 5.0 GHz, while the higher-resolution VLBI images at 8.4 and 15.4 GHz reveal a central component {\rm B} between the two lobes \citep[Figure \ref{fig:j1511}: the current paper; Paper II;][]{OD08}. {\rm A} and {\rm C} show convex spectra with a peak at about 8 and 5 GHz, respectively. {\rm B} exhibits a rising spectrum that probably breaks at $\gtrsim $15 GHz and identifies {\rm B} as the core of the CSO \citep[Paper II;][]{OD08}.  

J1511+0518 has been extensively observed at 15 GHz (the MOJAVE campaign). Figure \ref{fig:j1511} shows the 2-dimensional distribution of the VLBI components derived from the archival 15-GHz data during 2005 and 2009. 
The hot spot {\rm A} shows a marginal proper motion of $7.0\pm3.3$ $\mu$as yr$^{-1}$ (or $0.038\pm0.018\,c$) to the East. It is in agreement with the previous report of hotspot-hotspot expansion velocity of 0.1$c$ \cite{OD08}. 
The arm length {\rm C1-B} is not found to change significantly in the jet-axis direction, but the peak position of {\rm C1} drifts through the lobe. The proper motion of {\rm C2} along the jet axis {\bf is} $v_r = 0.15\pm0.02\,c$.  {\rm C2} seems to move away from the hot spot {\rm C1} and may represent a deflected outflow originating  from the primary hot spot {\rm C1} \cite{Lai81,LB86}. The kinematic age based on the proper motion of {\rm A} is $300\pm140$ year. 

\subsection{J1734+0926 (CSO)}

\begin{figure*}
\begin{center}
\includegraphics[scale=0.775]{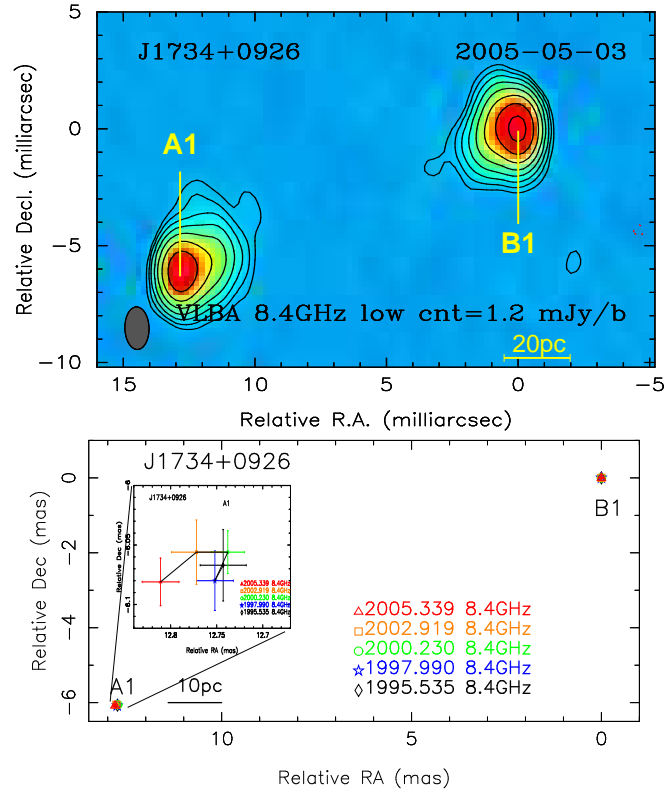}
\caption{The CSO source J1734+0926 with a missing core component. {\it Top}: The 8.4-GHz image of J1734+0926. {\it Bottom}: The 2D distribution of the VLBI components.}
\label{fig:j1734}
\end{center}
\end{figure*}

CSO source J1734+0926 (PKS 1732+094) is associated with a galaxy of magnitude $m_{R_C}=20.8$ and at redshift of 0.735 (1 mas = 7.086 pc) \cite{deV07}. The source is identified as a CSO based on the morphology with two edge-brightened hot spots 
($\alpha^{15.4}_{2.3}(A)= 1.1$, $\alpha^{15.4}_{2.3}(B)= 1.3$: Paper II) and an extension toward the geometric center, where the core is too weak to be detected \citep{Sta99,Pec00}. Figure \ref{fig:j1734}(top) displays the radio structure of J1734+0926. The projected separation between two hot spots is about 100 pc. The positional variation of the hot spot {\rm A1} relative to {\rm B1} is displayed in the inset of Figure \ref{fig:j1734}(bottom). The separation of {\rm A1-B1} shows a back-and-forth oscillation with an overall expansion rate of $6.3\pm3.0$ $\mu$as yr$^{-1}$ (or $0.12\pm0.05\,c$), corresponding to a kinematic age of $1300\pm620$ year.

\subsection{J1756+5748 (Core-Jet)}

\begin{figure*}
\begin{center}
\includegraphics[scale=1.0]{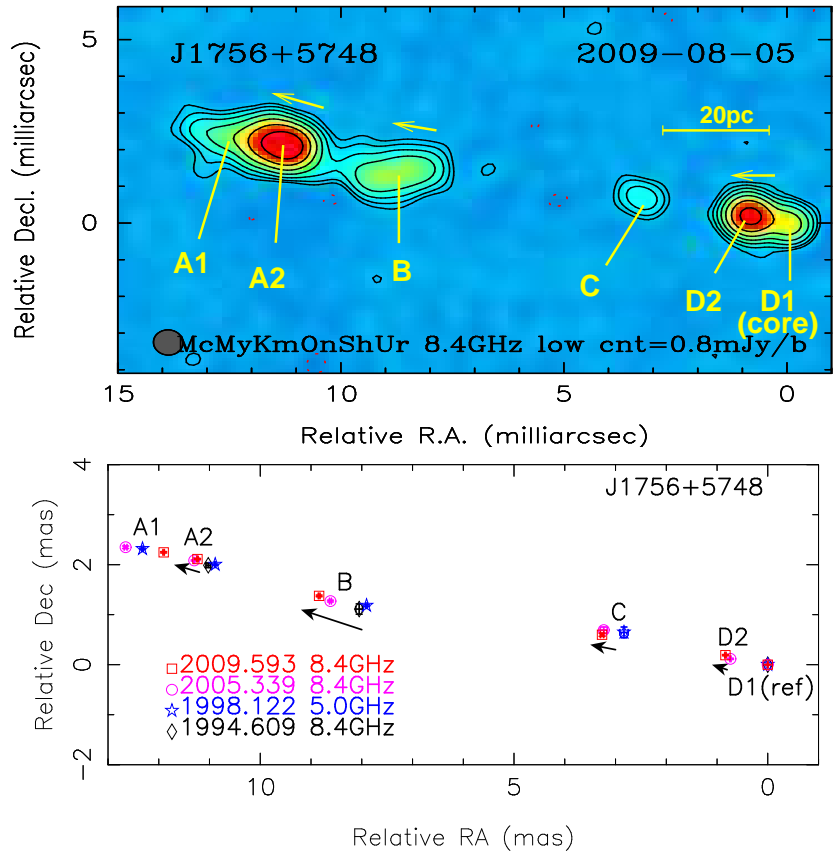}
\caption{The Core-Jet source J1756+5748 with core component {\rm D1}. {\it Top}: the CLEANed image of J1756+5748 at epoch 2009 August 5. {\it Bottom}: The 2-D distribution of the jet components.}
\label{fig:j1756}
\end{center}
\end{figure*}

\begin{figure*}
\begin{center}
\includegraphics[scale=0.75]{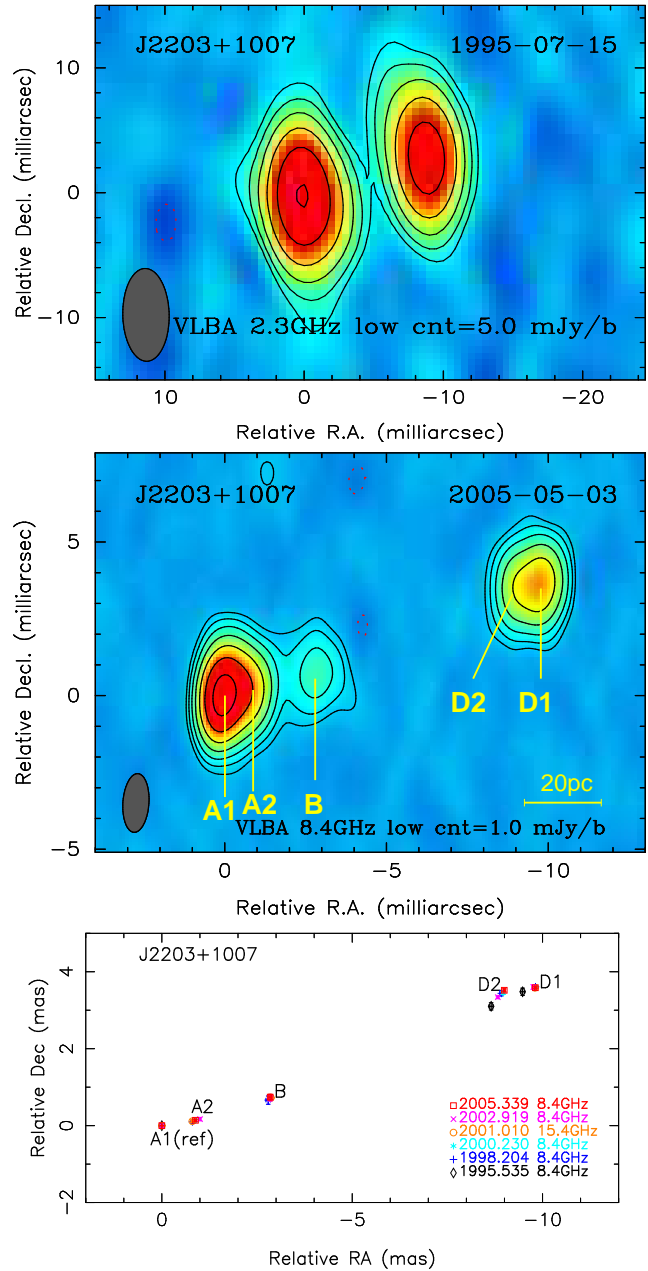}
\caption{The CSO source J2203+1007 with a missing central core component. {\it Top}: The 2.3-GHz image of J2203+1007. {\it Middle}: The 8.4-GHz image of J2203+1007. {\it Bottom}: the 2D distribution of the VLBI components.}
\label{fig:j2203}
\end{center}
\end{figure*}

\begin{figure*}
\begin{center}
\includegraphics[scale=0.75]{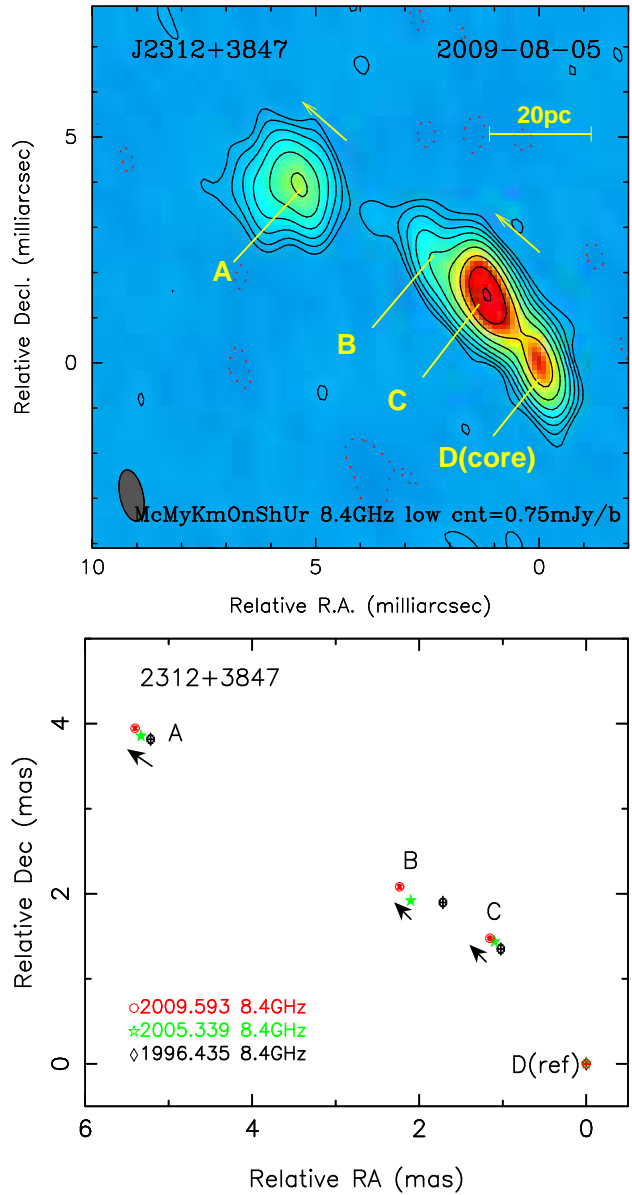}
\caption{The Core-Jet source J2312+3847 with radio core {\rm D}. {\it Top}: The 8.4-GHz CLEANed image of J2312+3847 at epoch 2009 August 5. {\it Bottom}: The 2-D distribution of the jet components.}
\label{fig:j2312}
\end{center}
\end{figure*}

Core-jet source J1756+5748 has been identified with a high-redshift quasar ($z=2.11$; Henstock et al. 1997; 1 mas = 8.201 pc) of magnitude 18. The top panel in Figure \ref{fig:j1756} shows the 8.4-GHz image of J1756+5748 obtained from the observations in 2009. The radio structure is characterized by a series of bright knots in the East-West direction, consistent with previous observations at 5 GHz \citep{Tay94,Bri08}. From the observations reported in Paper II, the knot {\rm D1} at the western extreme of the source has the flattest spectral index $\alpha^{15.4GHz}_{8.4GHz}=0.0\pm0.1$, and is identified as the radio core. The other components show steep spectra with $\alpha^{15.4}_{2.3}>0.7$. J1756+5748 is classified as a core-jet source.

Figure \ref{fig:j1756}(bottom) displays the 2-dimensional distribution of the jet components and their general eastern motion relative to {\rm D1}. The component {\rm A1} moves to the northeast from epoch 1998 to 2005, whereas it appears to move backwards from 2005 to 2009. We note that this apparent backward motion of {\rm A1} probably results from internal changes of the  emission structure in the eastern knots. {\rm A1} is located in the vicinity of the brightest VLBI component {\rm A2} and thus is prone to a change of the emission structure of {\rm A2}. The intensity change of the brightest jet knot {\rm A2} likely result in the apparent inward motion of {\rm A1}. A linear regression fit has been made to obtain the proper motion velocities of the components along the jet axis of $80\degr$:  for {\rm A2} 31.8$\pm$1.3 $\mu$as yr$^{-1}$ , for {\rm B} 82.5$\pm$1.8 $\mu$as yr$^{-1}$, and for {\rm C} $34.0\pm7.2$ $\mu$as yr$^{-1}$. These angular velocities correspond to $2.6\,c$, $6.9\,c$ and $2.8\,c$ for A2, B and C, respectively. In 1994 and 1998, the inner jet knot {\rm D2} was still blended with the core {\rm D1}. Since epoch 2005, {\rm D2} has been well separated from {\rm D1}. The 2005 and 2009 data give a proper motion of {\rm D2} of $\mu_r=$25.5 $\mu$as yr$^{-1}$ to the northeast. The kinematic ages of the jet components derived from the proper motions are $116\pm5$ year (A2), $35\pm1$ year (B), $31\pm7$ year (C) and $\sim$11 year (D2). The proper motion direction and the gradually increasing kinematic ages downstream in the jet confirm the core-jet identification of the source.

\subsection{J2203+1007 (CSO)}

The CSO source J2203+1007 is likely associated with a galaxy of magnitude 22.02 at a redshift of 1.005 (1 mas = 7.830 pc) \cite{Hea08}. The radio spectrum of the source shows a convex spectrum with a peak at $\sim$5 GHz, which identifies it as a HFP \cite{Dal00}. The VLBI images of the source display a double-component morphology at 1.7 and 2.3 GHz.
At 5.0 GHz and higher frequencies, both the eastern and western component shows a steep spectrum with $\alpha^{15.4}_{5.0}=1.1\pm0.2$ (East) and $\alpha^{15.4}_{5.0}=1.3\pm0.2$ (West). 
The morphology of both components (Figure \ref{fig:j2203}) display sharp outer edges and an extension toward the geometric center similar with that of J1734+0926. All the lobes and internal jet components are found to have steep spectral indices between 5.0 and 15.4 GHz ranging from 1.2 to 1.7 (Paper II). The central core probably lies between components {\rm B} and {\rm D2} but it is too weak to be detected. The western components {\rm D2} and {\rm D1} show relative proper motion of $21.1\pm4.8$ $\mu$as yr$^{-1}$ ($1.13\pm0.26\,c$: D2) and $10.3\pm3.6$ $\mu$as yr$^{-1}$ ($0.53\pm0.18\,c$: D1) with respect to {\rm A1}. 
The angular velocity of {\rm D1} is consistent with the upper limit derived by Gugliucci et al. (2005). 
The kinematic age for this source is about $500\pm180$ year.

\subsection{J2312+3847 (Core-Jet)}

Core-jet source J2312+3847 is a high-redshift ($z$=2.17; 1 mas = 8.171 pc) quasar with an optical magnitude of 17.5 \cite{Hew93}. The source exhibits a double-component structure in northeast-southwest direction at frequencies below 8 GHz, and the two components have almost equal flux densities (VCS archive). At 8.4 and 15.4 GHz, the southwest component is resolved into three sub-components, and the whole source displays a triple morphology \citep[Figure \ref{fig:j2312}: the current paper;][]{Hen95}. The radio structure of the source can be fitted with four components, namely {\rm A, B, C} and {\rm D} from northeast to southwest. {\rm D} has a flat spectrum with a spectral index $\alpha^{15.4GHz}_{8,4GHz}=-0.3$ and therefore is identified as the radio core (Paper II). The position angles of the jet components show an increasing trend from the innermost outwards, likely corresponding to a smooth jet bending. 

Linear regression fitting gives a proper motion of A, $29.5\pm4.6$ $\mu$as yr$^{-1}$ (or $2.5\pm0.4\,c$) along a position angle 54\degr. The kinematic age of {\rm A} derived from the proper motion is 71$\pm$16 year. 
Britzen et al. measured the expansion velocity of {\rm D} \citep[their definition 'C2':][]{Bri08} as $v_r=4.4\pm2.1\,c$ using component {\rm A} as a reference from 5-GHz VLBI data sets. This separation velocity of {\rm A-D} is consistent with ours within the uncertainty. The proper motion of {\rm B} is $61.3\pm4.7$ $\mu$as yr$^{-1}$ (or $5.2\pm0.4\,c$) along the jet axis, corresponding to a kinematic age of only 16$\pm$1 year. {\rm C} is the brightest  component in J2312+3847 and moves to the northeast with a velocity of $23.0\pm4.3$ $\mu$as yr$^{-1}$ (or $1.9\pm0.4\,c$). The derived kinematic age of $\rm C$ of 26$\pm$5 year is a bit higher than that of $\rm B$. A possibility for the lower velocity (or higher age) of {\rm C} is that {\rm C} has passed through a stationary stage during the observing sessions. Stationary components located at the starting section of the supersonic jet seem to be common in blazars at a distance of a few parsecs away from the core \cite{Jor01}. Such stationary features could correspond to standing re-collimation shocks, which result from abrupt changes in the pressure of the external interstellar medium, or are generated as a result of deflection by dense clouds. In the re-collimation shocks or compressed shocks at the interface of jet-cloud interaction, both the particle density and magnetic field energy density might be substantially enhanced, resulting in brightening of the stationary jet components. If the jet losses a significant fraction of the kinetic power at the stationary shocks, the jet fluid may not maintain laminar flow and becomes turbulent downstream \cite{KB07}. In FR I jets, bright hot spots are often observed at the transition (hundred of parsecs from the core) between collimated and diffuse jet  \citep[{\it e.g.}, hot spot B in 3C~48:][]{Wil91,Wor04,Feng05,An10}, indicating that jet kinetic energy is released through radiation at the standing shock. 

\section{Summary and Conclusion}

New 8.4-GHz observations have been presented for four sources from a subsample of ten CSO candidates with a VLBI network consisting of four Chinese and two European telescopes on 2009 August 5. The four sources have been mapped with high sensitivity with a typical noise of 0.3 mJy beam$^{-1}$ and with sub-milliarcsecond resolution. The source structures are in excellent agreement with previous VLBI observations. The observations demonstrate that the inclusion of the two new Chinese telescopes (Miyun 50-meter and Kunming 40-meter) greatly improves the ({\it u,v}) coverage of the EVN-CVN on long baselines and significantly increases the detectability of the VLBI network to map fine structures of compact astronomical objects. 

Use has also been made of earlier VLBA observational data in 2005 and archival data of ten CSO candidates to determine the expansion velocities of the hot spots. The long time baseline between the present observations and the earliest ones resulted in a better determination and constraint of the expansion velocities of hot spots and proper motions of the jet components with a highest accuracy of $\sim$1.3 $\mu$as yr$^{-1}$. 

The major results from the current paper can be summarized as follows : 
\newline 
(1) 
Six out of ten sources are identified as CSOs, and the expansion velocities or limits are determined for all ten sources. 
Two of these CSOs (J1734+0926 and J2203+1007) are characterized by symmetric double hot spots with sharp leading edges and extensions toward the geometric center in the 8.4-GHz images, while the central cores are not detected at this frequency. The remaining four CSOs (J0132+5620, J0518+4730, J1335+5844, J1511+0518) show triple morphology with a central core and two lobes at about equal separations. J1324+4048 is identified as a CSO candidate, and further high-resolution observations are needed to clarify its nature. Two core-jet sources (J1756+5748 and J2312+3847) display one-sided knotty jets with the core at the end of the jet. J0017+5312 is likely a core-jet source, but an unambiguous CSO identification requires the detection of the central core from much higher-resolution observations.    
  
Relative proper motions of the VLBI components in all ten sources have been measured or constrained. For the CSOs with known redshift, the kinematic ages in the source frame range from 300 to 1800 years. The core-jet sources show apparent superluminal motions, suggesting relativistic beaming of the jet flow in these sources.  
 
The internal jet components in the CSO J0132+5620 shows mirror symmetry relative to the central position, confirming the CSO identification of the source. The proper motion velocities of the internal jet knots decrease with increasing distance from the core, likely reflecting the density gradient of the ISM surrounding in the host galaxy. The evolution of the jet velocity with time or radial distance is an important ingredient for sophisticated models describing the dynamical evolution of CSOs and the physical environment in the innermost host galaxies.

\noindent
(2) 
The present work adds seven new CSOs with proper motion measurements to the CSO proper motion sample \citep[the current paper;][]{Tay00,Pol03,Gir03,Gug05,Nag06,GP09}. The number of CSOs with known expansion velocities (or limits) has grown from 30 to 37, an increase of 23\%. Among the CSO proper motion sample, 27 sources have redshifts and the source-frame kinematics ages. Eleven out of these 27 have kinematic ages less than 500 years, indicating an overabundance of the young CSOs, which confirms earlier results \cite{Gug05}. 
An increasing sample of CSOs with known proper motions will be crucial for clarifying their physical nature and for understanding the dynamical evolution of this class of compact radio sources. 

\noindent
(3)  
Non-radial motion appears not to be unusual in CSOs. Transverse motions or jet head wandering have been found in both  young and intermediate-age CSOs. The hot spots or lobes manifesting transverse motions are also found morphology distortion. Examples of such motions are found in the reflection of the western jet in J1511+0518, and the elongated trail perpendicular to jet axis in J1335+5844. A variety of mechanisms initially proposed for FR II lobes may also be responsible for these complex motion patterns of CSO hot spots, such as changes of the beam direction \cite{Sch82}, jet deflections \cite{LB86}, collimated outflow originating from the primary hot spot \cite{Lai81}, and others. In theoretical models of CSO evolution, a necessary requirement for a CSO to grow into an FR II galaxy is that the jet remains stable against the turbulence before escaping the host galaxy \cite{KA97, KB07}. In other words, the CSOs with distorted jet heads (hot spots) may develop turbulent flow during the CSO or CSS phase, and accordingly these sources will evolve into FR I sources. On the other hand, the CSOs with compact and well-defined lobes characterized by sharp edge-brightened interfaces maintain laminar jet flow and have a chance to evolve into FR IIs. The detailed discussion of the dynamic evolution of CSOs is presented in An \& Baan (in prep.).

\section*{Acknowledgement}

We would like to thank the anonymous referee for his/her constructive comments which help to clarify the paper.
The authors are grateful to the CVN and EVN staff for facilitating these observations, and to the JIVE staff for correlating the data and for help with coordinating the observations. T.A. thanks the Overseas Research Plan for CAS-Sponsored Scholars, the Netherlands Foundation for Sciences (NWO) and the China-Hungary Collaboration and Exchange Program of the CAS. T.A. thanks Carlo Stanghellini providing the epochs 2004 and 2006 data of J1335+5844 and Daniele Dallacasa for discussion of the CSO identification of J1335+5844. F.W. thanks the JIVE Summer Student Program. L.C. thanks the the program of the Light in China's Western Region (Grant No. XBBS201024). The European VLBI Network is a joint facility of European, Chinese, South African and other radio astronomy institutes funded by their national research councils. The National Radio Astronomy Observatory is a facility of the National Science Foundation operated under a cooperative agreement by Associated Universities, Inc. This research has made use of data from the MOJAVE database that is maintained by the MOJAVE team (Lister et al., 2009, AJ, 137, 3718). This research has made use of the NASA/IPAC Extragalactic Database (NED), which is operated by the Jet Propulsion Laboratory, California Institute of Technology, under contract with the National Aeronautics and Space Administration. This work is supported in part by the National Natural Science Foundation of Science and Technology of China (Grant No. 2009CB24900), the Science \& Technology Commission of Shanghai Municipality (06DZ22101), and the Knowledge Innovation Program of the Chinese Academy of Sciences.


\begin{table}[htbp]
    \begin{center}
\caption{Summary of Source Properties}
\label{tab:sample}
      \begin{tabular}{lcccccc}
    \hline
    \hline
Source & R.A. (J2000) & Dec. (J2000) & ID & $m$ & $z$ & Ref.  \\
(1)  & (2) & (3) & (4) & (5) & (6) & (7)  \\ 
\hline
J0017+5312 & 00 17 51.760 & +53 12 19.121 & QSO& 18.3  & 2.574 & S05 \\
J0132+5620$^*$ & 01 32 20.447 & +56 20 40.370 &    &       &       &  \\
J0518+4730 & 05 18 12.090 & +47 30 55.528 &    &       &       &     \\
J1324+4048$^*$ & 13 24 12.096 & +40 48.11.763 & QSO& 19.5  & 0.496 & V96\\
J1335+5844 & 13 35 25.928 & +58 44 00.291 &    & 22.0  & 0.57 & S09\\
J1511+0518 & 15 11 41.266 & +05 18 09.259 & G  & 17.2  & 0.084 & C01\\
J1734+0926 & 17 34 58.377 & +09 26 58.260 & G  & 20.8  & 0.735 & D07\\
J1756+5748$^*$ & 17 56 03.628 & +57 48 47.998 & QSO& 18.0  & 2.110 & H97\\
J2203+1007 & 22 03 30.953 & +10 07 42.589 & G  & 22.02 & 1.005 & H08\\
J2312+3847$^*$ & 23 12 58.794 & +38 47 42.661 & QSO& 17.5  & 2.170 & HB89\\
\hline
\end{tabular}
  \end{center}
Column (1) lists the sources discussed in the current paper. The source names with $^*$ are included in the 2009 observations.
Columns (2) and (3) list the coordinates (J2000) of the pointing centers used in the VLBI observations.
Column (4) gives the optical classification of the host galaxy. 
Columns (5) and (6) present the magnitude and redshift of the host galaxies of the observed radio sources. 
References: S05 - Sowards-Emmerd et al. 2005; V96 -  Vermeulen et al. 1996; S09 - Stanghellini et al. 2009; C01 - Chavushyan et al. 2001; D07 - de Vries et al. 2007; H97 -  Henstock et al. 1997; H08 - Healey et al. 2008; HB89 - Hewitt \& Burbidge 1989.
\end{table}

\clearpage


\begin{landscape}
\begin{table}[htbp]
\centering
 \renewcommand\tabcolsep{4pt}
\caption{Observational and Image Parameters}
 \begin{tabular}[t]{lllcccccl}
\hline\hline
Source & Fig. &Date &$\nu$ & $\tau$ & Restoring Beam & $I_{peak}$ & $\sigma_{rms}$ &  Contours\\ 
       &      &     & (GHz)&(hour) & (mas,mas,deg)  & (mJy b$^{-1}$)    & (mJy b$^{-1}$)   & (mJy b$^{-1}$)   \\
\hline
J0017+5312&\ref{fig:j0017}-a &2005 MAY 3$^a$ &8.4 &0.8 & 1.30$\times$0.80, 0.0   & 242 & 0.42  & 2.0$\times$(-1,1,2,4,...64) \\
          &\ref{fig:j0017}-a &1994 AUG 12$^b$&8.3 &0.1 & 1.30$\times$0.80, 0.0   & 238 & 0.98  & 2.0$\times$(-1,1,2,4,...64) \\
J0132+5620&\ref{fig:j0132}-a &2009 AUG 5$^a$ &8.4 &4.0 & 0.76$\times$0.62, 21.6  & 199 & 0.25  & 0.75$\times$(-1,1,2,4,...256) \\
          &\ref{fig:j0132}-b &2005 MAY 3$^a$ &8.4 &0.8 & 1.21$\times$0.84,$-$39.5& 179 & 0.33  & 1.00$\times$(-1,1,2,4,...128) \\
          &\ref{fig:j0132}-c &2002 DEC 2$^b$ &8.4 &1.1 & 1.60$\times$1.00, 0.0   & 204 & 0.15  & 0.45$\times$(-1,1,2,4,...256) \\
          &\ref{fig:j0132}-d &2000 MAR 25$^b$&8.4 &1.1 & 1.60$\times$1.00, 0.0   & 214 & 0.15  & 0.45$\times$(-1,1,2,4,...256) \\
J0518+4730&\ref{fig:j0518}   &2005 MAY 3$^a$ &8.4 &0.8 & 1.35$\times$0.78,$-$9.9 & 131 & 0.42  & 1.30$\times$(-1,1,2,4,...64)\\
J1324+4048&\ref{fig:j1324}-a &2009 AUG 5$^a$ &8.4 &3.6 & 1.01$\times$0.63, 29.9  & 121 & 0.25  & 0.75$\times$(-1,1,2,4,...128) \\
J1335+5844&\ref{fig:j1335}   &2004 JAN 25$^c$&8.4 &3.0 & 1.72$\times$0.91, 24.8 & 351 & 0.08& 0.25$\times$(-1,1,2,4,...256) \\
J1511+0518&\ref{fig:j1511}   &2005 MAY 3$^a$ &8.4 &0.6 & 1.05$\times$0.47, $-$0.7& 213 & 0.40& 1.20$\times$(-1,1,2,4,...128) \\ 
J1734+0926&\ref{fig:j1734}   &2005 MAY 3$^a$ &8.4 &0.8 & 1.82$\times$0.91, 1.2 & 197 & 0.40& 1.2$\times$(-1,1,2,4,...128) \\
J1756+5748&\ref{fig:j1756}   &2009 AUG 5$^a$ &8.4 &3.6 & 0.69$\times$0.66, 13.7  &  97 & 0.27  & 0.80$\times$(-1,1,2,4,...64)  \\
J2203+1007&\ref{fig:j2203}   &2005 MAY 3$^a$ &8.4 & 0.8& 1.91$\times$0.80, $-$3.6&  87 & 0.30  & 1.00$\times$(-1,1,2,4,...64)  \\
          &                  &1995 JUL 15$^b$&2.3 &0.1 & 7.43$\times$3.32, 1.0   & 166 & 1.70  & 5.00$\times$(-1,1,2,4,...32) \\
J2312+3847&\ref{fig:j2312}   &2009 AUG 5$^a$ &8.4 &3.1 & 1.15$\times$0.52, 12.4  & 200 & 0.25  & 0.75$\times$(-1,1,2,4,...256) \\
\hline
 \end{tabular}\\
Notes: (a) The current paper;
(b) The archive of the VLBA Calibrator Survey;
(c) The original image is referred to Stanghellini et al. (2009). 

\label{tab:map}
\end{table}
\end{landscape}

\clearpage
{\small 
\begin{longtable}[c]{l*{9}{c}}
\caption{Model fitting parameters}  \\[2mm]
\hline
Epoch  & $\nu$ & Comp. &ID &  R  & $\theta$ & S$_{int}$ &$\theta_{maj}$ & $\theta_{min}$ & P.A.  \\
       &       & (GHz) &   &(mas)& (deg)    &(mJy)       & (mas)         & (mas)        & (\degr) \\
(1)    &  (2)  & (3)   &(4)& (5) & (6)      &(7)        &(8)            &(9)           &(10)     \\
\hline\hline 
\endhead
\hline
\multicolumn{10}{r}{\small \slshape continued overleaf} 
\endfoot
\hline\\[-6pt]
\endlastfoot
\multicolumn{10}{c}{J0017+5312 (Core-Jet)} \\
2005.336 &8.4 & A &j& 0.0         & 0.0        &298.1(17.9)& 0.62  & 0.18  &$-$59.8 \\
         &    & B &j& 1.777(0.015)&$-$64.8(0.2)&246.9(14.9)& 0.44  & 0.19  &$-$69.1 \\
         &    & C &j& 5.073(0.088)&$-$65.0(0.5)&9.7(2.6)   & 1.20  & 1.20  &        \\
1994.609 &8.4 & A &j& 0.0         & 0.0        &293.5(17.7)& 0.65  & 0.13  &$-$60.8 \\
         &    & B &j& 1.909(0.034)&$-$65.2(0.4)&226.2(13.8)& 0.84  & 0.30  &$-$66.6 \\
         &    & C &j& 5.440(0.210)&$-$63.8(1.3)& 12.2(4.1) & 1.06  & 1.06  &        \\
\hline
\multicolumn{10}{c}{J0132+5620 (CSO)} \\
2009.593 &8.4 & D1&h&12.181(0.028)&$-$77.6(0.1)&298.5(17.9)& 0.62  & 0.37  & 31.6   \\
            & & D2&j&11.547(0.029)&$-$80.8(0.1)& 37.3(2.3) &$<$0.15&$<$0.15&        \\
            & & C &j&10.539(0.044)&$-$79.4(0.1)& 21.3(1.6) & 1.42  & 1.42  &        \\
            & & B1&j& 2.383(0.107)&$-$85.5(1.8)&  5.5(0.8) & 1.03  & 1.00  &        \\
            & & A &h&0.0          &  0.0       & 39.1(2.4) & 0.78  & 0.40  & 62.3   \\
2005.336 &8.4 & D1&h&12.093(0.014)&$-$77.2(0.1)&227.0(13.7)& 0.64  & 0.38  & 21.8  \\
            & & D2&j&11.326(0.029)&$-$80.9(0.1)& 22.1(1.5) &$<$0.20&$<$0.20&       \\
            & & A &h&0.0          &0.0         & 36.3(2.4) & 1.03  & 0.45  & 75.5  \\ 
2002.919 &8.4 & D1&h&12.108(0.034)&$-$77.2(0.1)&233.1(14.0)& 0.68  & 0.35  &24.3   \\
            & & D2&j&11.260(0.036)&$-$80.7(0.1)& 21.9(1.4) &$<$0.10&$<$0.10&       \\
            & & C &j&10.058(0.057)&$-$79.7(0.2)&  9.0(0.6) & 1.48  & 1.48  &       \\
            & & B2&c& 5.491(0.137)&$-$80.7(1.0)& 2.3(0.3) &$<$1.03&$<$1.03&       \\
            & & B1&j& 2.583(0.076)&$-$87.3(1.3)&  5.7(0.5) & 1.72  & 1.72  &       \\
            & & A &h& 0.0         & 0.0        & 36.7(2.2) & 0.97  & 0.43  &72.9   \\
2001.010 &15.3& D1&h&12.126(0.037)&$-$77.2(0.1)& 92.4(6.5) & 0.54  & 0.27  & 32.7  \\
            & & D2&j&11.204(0.041)&$-$80.5(0.1)& 14.4(1.1) & 0.56  & 0.56  &       \\
            & & A &h&0.0          & 0.0        &  7.8(0.7) & 0.98  & 0.45  & 77.6  \\
2000.230 &8.4 & D1&h&12.127(0.033)&$-$77.3(0.1)&243.3(14.6)& 0.65  & 0.32  &31.1   \\
            & & D2&j&11.265(0.034)&$-$80.7(0.1)& 22.8(1.4) &$<$0.25&$<$0.25&       \\
            & & C &j&10.164(0.045)&$-$79.9(0.2)& 11.2(0.7) & 1.53  & 1.53  &       \\
            & & B2&c& 5.505(0.101)&$-$80.6(0.8)& 1.9(0.2)  &$<$0.45&$<$0.45&       \\
            & & B1&j& 3.148(0.067)&$-$84.0(0.9)&  3.6(0.3) & 0.74  & 0.74  &       \\
            & & A &h& 0.0         &0.0         & 39.4(2.4) & 1.01  & 0.44  &74.1   \\
1998.204 &8.4 & D1&h&12.110(0.019)&$-$77.4(0.1)&226.0(13.6)& 0.58  & 0.31  &29.4   \\
            & & D2&j&11.180(0.032)&$-$81.6(0.1)& 15.4(1.2) &$<$0.05&$<$0.05&       \\
            & & C &j& 9.821(0.095)&$-$81.1(0.4)&  8.1(0.9) & 1.16  & 1.16  &       \\
            & & A &h& 0.0         &0.0         & 36.2(2.3) & 0.97  & 0.63  &48.1   \\
1998.204 &5.0 & D1&h&12.055(0.015)&$-$76.8(0.1)&385.6(19.3)& 0.79  & 0.20  &18.7   \\
            & & D2&j&11.086(0.024)&$-$80.4(0.1)& 46.3(2.4) & 0.43  & 0.43  &       \\
            & & C &j& 9.000(0.142)&$-$77.6(0.6)&  9.5(0.9) & 1.77  & 1.77  &       \\
            & & B1&j& 3.040(0.070)&$-$84.0(1.0)& 19.4(1.3) & 2.14  & 2.14  &       \\
            & & A &h& 0.0         &  0.0       &140.8(7.1) & 0.87  & 0.59  &76.3   \\
1994.609 &8.3 & D1&h&12.147(0.015)&$-$77.4(0.1)&276.2(13.9)& 0.64 & 0.26 & 36.0 \\
            & & D2&j&11.162(0.053)&$-$80.8(0.2)& 38.4(2.9) & 0.69 & 0.69 &      \\
            & & A &h& 0.0         & 0.0        & 39.5(2.9) & 0.77 & 0.56 & 89.1 \\ 
\hline
\multicolumn{10}{c}{J0518+4730 (CSO)} \\
2005.336 &15.4& A &h& 1.939(0.026)&131.4(0.2)  &17.9(1.4)  & 0.55 & 0.30 &$-$54.6\\
         &    & B &c& 0.0         & 0.0        &99.6(7.0)  & 0.36 & 0.19 &77.9   \\
         &    & C &h& 3.004(0.017)&$-$66.1(0.1)&60.0(4.3)  & 1.11 & 0.52 &$-$75.4\\
         & 8.4& A &h& 1.921(0.022)&131.6(0.1)  &42.4(2.7)  & 0.76 & 0.40 &$-$50.5\\
         &    & B &c& 0.0         &0.0         &161.6(9.7) & 0.53 & 0.42 &$-$46.2\\
         &    & C &h& 3.054(0.016)&$-$66.2(0.1)&146.4(8.8) & 1.32 & 0.87 &$-$73.2\\
2002.919 & 8.4& A &h& 1.893(0.019)&130.5(0.1)  &41.8(2.5)  & 0.73 & 0.37 &$-$33.9\\
         &    & B &c& 0.0         &0.0         &160.6(9.6) & 0.67 & 0.43 &$-$41.7\\
         &    & C &h& 3.147(0.015)&$-$65.9(0.1)&165.1(10.0)& 1.59 & 0.79 &$-$78.6\\
2001.010 &15.4& A &h& 1.905(0.042)&130.6(0.4)  &18.2(1.7)  & 0.56 & 0.37 &$-$26.3\\
         &    & B &c& 0.0         &0.0         &88.9(6.3)  & 0.29 & 0.20 &$-$57.2\\
         &    & C &h& 3.155(0.027)&$-$65.7(0.3)&68.0(5.1)  & 1.54 & 0.61 &$-$83.7\\
2000.230 & 8.4& A &h& 1.782(0.018)&131.5(0.1)  &40.5(2.5)  & 0.78 & 0.48 &$-$53.3\\
         &    & B &c&  0.0        & 0.0        &149.9(9.0) & 0.57 & 0.29 &$-$45.2\\
         &    & C &h& 3.184(0.015)&$-$65.6(0.1)&174.1(10.5)& 1.47 & 0.76 &$-$76.9\\
1998.204 & 8.4& A &h& 1.758(0.029)&132.6(0.2)  &32.8(2.0)  & 0.39 & 0.39 &       \\
         &    & B &c& 0.0         &0.0         &137.7(8.3) & 0.53 & 0.21 &$-$51.4\\
         &    & C &h& 3.250(0.015)&$-$65.3(0.1)&171.7(10.3)& 1.32 & 0.60 &$-$78.3\\
1996.608 & 8.4& A &h& 1.606(0.068)&134.5(0.5)  &36.2(2.5)  & 0.89 & 0.89 &       \\
         &    & B &c& 0.0         &0.0         &111.6(6.8) & 0.28 & 0.28 &       \\
         &    & C &h& 3.281(0.049)&$-$65.9(0.3)&152.7(9.3) & 1.01 & 1.01 &       \\
\hline
\multicolumn{10}{c}{J1324+4048 (CSO Candidate)} \\
2009.593 & 8.4& A &h& 0.0         & 0.0         &146.1(8.9) & 0.65 & 0.39 &78.1   \\
            & & B2&j& 4.884(0.039)&$-$84.5(0.3) & 11.8(1.1) & 0.1  & 0.1  &       \\
            & & B &h& 5.504(0.024)&$-$83.0(0.2) &149.8(9.0) & 0.48 & 0.31 &6.3    \\
2005.336 & 8.4& A &h& 0.0         & 0.0         & 102.5(6.2)& 0.68 & 0.33 &77.3   \\
            & & B2&j& 4.561(0.066)&$-$84.0(0.6) &   6.9(0.7)& 0.1  & 0.1  &       \\
            & & B &h& 5.439(0.015)&$-$82.7(0.1) & 107.0(6.5)& 0.44 & 0.37 &29.5   \\ 
1998.122 & 5.0& A &h& 0.0         & 0.0         &187.6(9.4) & 0.68 & 0.47 & 73.8  \\
            & & B2&j& 4.351(0.046)&$-$82.5(0.4) & 10.4(0.6) & 0.1  & 0.1  & 173.7 \\
            & & B &h& 5.409(0.030)&$-$82.7(0.2) &161.2(8.1) & 0.58 & 0.37 & 24.8  \\ 
1996.433 &8.3 & A &h& 0.0         & 0.0         &130.4(7.9) & 0.59 & 0.42 & 66.2  \\
            & & B &h& 5.424(0.018)&$-$82.9(0.1) &128.5(7.8) & 0.77 & 0.52 &$-$0.3 \\
1993.446$^a$& 5.0& A &h& 0.0         & 0.0         &226.0      & 0.62 & 0.58 &$-$90.9\\ 
            & & B &h& 5.42        &$-$82.5      &187.0      & 0.74 & 0.63 & 20.1 \\   
\hline
\multicolumn{10}{c}{J1335+5844 (CSO)} \\
2006.839 & 8.4&A1&h&0.0           &0.0          &371.8(22.3)&0.50   & 0.27 &$-$21.7\\
         &    &A2&j&0.664(0.046)  &117.2(1.1)   & 40.6(2.5) &0.3    & 0.3  &       \\
         &    &A3&j&0.940(0.047)  &168.0(1.6)   & 22.6(1.4) &0.5    & 0.5  &       \\
         &    &A4&j&2.547(0.048)  &$-$166.6(0.9)& 14.3(0.9) &0.53   & 0.53 &       \\
         &    & C&c&5.616(0.116)  &$-$166.7(0.9)&  2.2(0.4) &0.54   & 0.54 &       \\
         &    &B3&j&11.573(0.057) &$-$169.1(0.2)&  9.7(0.9) &0.95   & 0.95 &       \\
         &    &B2&h&12.941(0.046) &$-$160.6(0.2)& 53.0(3.2) &0.95   & 0.95 &       \\ 
         &    &B1&h&13.000(0.046) &$-$164.7(0.2)& 99.2(6.0) &1.14   & 0.80 &$-$6.9 \\ 
2005.336 &15.4&A1&h&0.0           &0.0          &355.2(24.9)&0.29   & 0.13 &$-$33.6\\
         &    &A2&j&0.384(0.027)  &   138.0(0.3)& 55.2(5.1) &0.93   & 0.93 &       \\   
         &    &B1&h&13.012(0.032) &$-$164.2(0.1)& 43.3(4.8) &1.23   & 0.80 &$-$56.7\\   
2005.336 & 8.4&A1&h&0.0           &0.0          &358.9(21.5)&0.51   & 0.27 &$-$32.7\\
         &    &A2&j&0.522(0.040)  &144.4(0.8)   & 87.1(5.4) &1.04   & 1.04 &       \\
         &    &A3&j&2.595(0.078)  &$-$168.1(1.6)& 12.7(1.2) &0.71   & 0.71 &       \\
         &    &B3&j&11.158(0.108) &$-$170.0(0.5)&  7.3(1.0) &0.70   & 0.70 &       \\
         &    &B2&h&12.875(0.045) &$-$159.8(0.2)& 33.8(2.2) &0.74   & 0.74 &       \\ 
         &    &B1&h&12.984(0.039) &$-$164.2(0.1)&124.6(7.6) &1.11   & 1.11 &       \\ 
2004.063 & 8.4&A1&h&0.0           &0.0          &372.5(22.4)&0.50   & 0.26 &$-$21.1\\
         &    &A2&j&0.675(0.040)  &122.4(0.7)   & 48.8(3.0) &0.3    & 0.3  &       \\
         &    &A3&j&1.182(0.042)  &177.9(1.5)   & 17.2(1.1) &0.5    & 0.5  &       \\
         &    &A4&j&2.725(0.048)  &$-$166.0(0.9)& 10.0(0.7) &0.45   & 0.45 &       \\
         &    & C&c&5.512(0.074)  &$-$168.0(0.6)&  2.1(0.4) &0.35   & 0.35 &       \\
         &    &B3&j&11.226(0.057) &$-$169.5(0.2)&  7.3(0.7) &0.68   & 0.68 &       \\
         &    &B2&h&12.827(0.045) &$-$160.2(0.2)& 43.6(2.7) &0.99   & 0.99 &       \\ 
         &    &B1&h&12.992(0.040) &$-$164.5(0.2)&112.0(6.8) &1.01   & 1.01 &       \\ 
2002.029 & 8.4&A1&h&0.0           &0.0          &328.6(19.7)&0.46   & 0.0  &$-$17.2\\
         &    &A2&j&0.572(0.068)  &121.8(1.4)   & 50.7(3.2) &0.47   & 0.47 &       \\
         &    &A3&j&0.945(0.067)  &174.5(2.7)   & 26.9(1.9) &0.46   & 0.46 &       \\
         &    &A4&j&2.298(0.071)  &$-$168.3(1.5)&  5.7(0.9) &0.33   & 0.33 &       \\
         &    &B3&j&11.135(0.129) &$-$170.0(0.5)&  6.2(1.1) &0.55   & 0.55 &       \\
         &    &B2&h&12.802(0.075) &$-$159.8(0.3)& 27.3(2.1) &0.85   & 0.85 &       \\ 
         &    &B1&h&12.971(0.068) &$-$164.1(0.3)&119.1(7.3) &0.98   & 0.98 &       \\ 
1994.609 & 8.4&A1&h&0.0           &0.0          &397.6(23.9)& 0.38  & 0.38 &       \\
         &    &A2&j& 0.938(0.042) &147.2(0.5)   & 42.4(2.7) & 0.3   & 0.3  &       \\
         &    &B2&h&12.754(0.057) &$-$158.8(0.2)& 21.0(1.7) & 0.65  & 0.65 &       \\
         &    &B1&h&12.940(0.039) &$-$163.7(0.2)&150.8(9.2) & 1.06  & 1.06 &       \\
\hline
\multicolumn{10}{c}{J1511+0518 (CSO)} \\
2009.405 &15.4& A &h& 2.202(0.016)&89.1(0.3)    &89.2(6.3)  & 0.41  & 0.25 &87.9   \\
         &    & B &c& 0.0         &0.0          &50.5(3.7)  & 0.23  & 0.23 &       \\
         &    & C1&h& 2.651(0.014)&$-$86.2(0.2) &340.6(23.9)& 0.29  & 0.10 &77.1   \\
         &    & C2&j& 3.454(0.018)&$-$93.2(0.3) &58.1(4.2)  & 0.45  & 0.18 &29.4   \\
2007.644 &15.4& A &h& 2.135(0.014)&89.4(0.3)    &171.1(12.0)& 0.41  & 0.26 &81.7   \\
         &    & B &c& 0.0         &0.0          &35.5(2.6)  & 0.26  & 0.26 &       \\
         &    & C1&h& 2.699(0.014)&$-$86.7(0.2) &365.6(25.6)& 0.28  & 0.15 &67.0   \\
         &    & C2&j& 3.440(0.016)&$-$92.3(0.2) &71.3(5.0)  & 0.42  & 0.20 &38.5   \\
2007.236 &15.4& A &h& 2.164(0.021)&89.1(0.4)    &177.8(12.5)& 0.41  & 0.25 &$-$81.1\\
         &    & B &c& 0.0         &0.0          &32.2(2.4)  & 0.29  & 0.29 &       \\
         &    & C1&h& 2.671(0.021)&$-$86.3(0.3) &348.2(24.4)& 0.28  & 0.06 &73.7   \\
         &    & C2&j& 3.407(0.022)&$-$91.9(0.3) &66.5(4.7)  & 0.44  & 0.21 &17.9   \\
2006.916 &15.4& A &h& 2.180(0.024)& 89.7(0.5)   &171.9(12.2)& 0.43  & 0.21 &85.9   \\
         &    & B &c&   0.0       & 0.0         & 24.0(2.6) & 0.40  & 0.40 &       \\
         &    & C1&h& 2.653(0.023)&$-$87.1(0.3) &342.5(24.0)& 0.27  & 0.16 & 66.2  \\
         &    & C2&j& 3.368(0.030)&$-$92.1(0.4) & 58.7(4.5) & 0.36  & 0.15 & 50.4  \\
2006.259 &15.4& A &h& 2.161(0.015)&89.4(0.3)    &165.8(11.7)& 0.37  & 0.31 &73.6   \\
         &    & B &c& 0.0         & 0.0         & 26.0(2.3) & 0.28  & 0.28 &       \\
         &    & C1&h& 2.676(0.014)&$-$86.9(0.2) &287.6(20.2)& 0.25  &0.06  &81.6   \\
         &    & C2&j& 3.353(0.021)&$-$91.7(0.3) & 50.4(3.8) & 0.44  &0.27  &54.9   \\
2005.336 &15.4& A &h& 2.163(0.015)&89.5(0.3)    &178.9(12.6)& 0.31  &0.16  &89.7   \\
         &    & B &c& 0.0         & 0.0         & 28.0(2.2) & 0.24  &0.24  &       \\
         &    & C1&h& 2.674(0.014)&$-$87.3(0.2) &243.7(17.1)& 0.28  &0.04  &71.5   \\ 
         &    & C2&j& 3.360(0.020)&$-$92.3(0.2) & 42.7(3.2) & 0.31  &0.19  &11.8   \\
2005.336 &8.4 & A &h& 2.178(0.068)&89.2(0.1)    &261.9(15.7)& 0.32  &0.29  &$-$46.4\\
         &    & B &c& 0.0         &0.0          & 18.5(1.2) & 0.38  &0.38  &       \\
         &    & C1&h& 2.710(0.060)&$-$87.7(0.5) &375.4(22.5)& 0.70  &0.31  &79.0   \\
         &    & C2&j& 3.378(0.060)&$-$93.4(0.5) & 80.3(4.8) & 0.43  &0.04  &$-$5.1 \\
\hline
\multicolumn{10}{c}{J1734+0926 (CSO)} \\
2005.336 & 8.4&B1 &h& 0.0         & 0.0         &219.4(13.2)& 0.52  & 0.35 &$-$51.7\\
         &    &B2 &j& 1.222(0.028)&115.4(0.4)   & 35.1(2.1) &$<0.05$&$<0.05$&      \\
         &    &A2 &j&13.330(0.028)&115.9(0.1)   & 26.1(1.6) &$<0.05$&$<0.05$&      \\
         &    &A1 &h&14.181(0.028)&115.4(0.1)   &166.8(10.0)& 0.62  & 0.29 &$-$63.3\\
2002.919 & 8.4&B1 &h& 0.0         & 0.0         &229.3(13.8)& 0.78  & 0.46 &$-$28.0\\
         &    &B2 &j& 0.888(0.039)&81.8(2.1)    & 71.5(4.3) &$<0.05$&$<0.05$&      \\
         &    &A2 &j&12.922(0.056)&118.8(0.1)   & 19.2(1.2) &$<0.05$&$<0.05$&      \\
         &    &A1 &h&14.135(0.038)&115.4(0.1)   &178.8(10.7)& 0.84  & 0.57 &$-$28.5\\
2000.230 & 8.4&B1 &h& 0.0         & 0.0         &159.3(9.6) & 0.67  & 0.47 &$-$32.6\\
         &    &B2 &j& 1.194(0.026)&114.6(0.5)   & 37.5(2.3) &$<0.05$&$<0.05$&      \\
         &    &A2 &j&12.799(0.027)&115.6(0.1)   & 32.3(2.0) &$<0.05$&$<0.05$&      \\
         &    &A1 &h&14.104(0.025)&115.4(0.1)   &159.3(9.6) & 0.81  & 0.64 &$-$49.5\\
1997.990 &15.4&B1 &h& 0.0         & 0.0         &70.3(4.9)  & 0.48  & 0.10 &$-$40.0\\
         &    &B2 &j& 0.978(0.052)& 87.8(2.4)   &28.3(2.1)  & 1.87  & 0.69 &$-$7.7 \\
         &    &A2 &j&12.870(0.051)&116.7(0.1)   &15.7(1.2)  & 1.87  & 0.07 & 26.9  \\
         &    &A1 &h&14.331(0.040)&115.1(0.1)   &52.2(3.7)  & 0.56  & 0.42 &$-$19.6\\
         & 8.4&B1 &h& 0.0         & 0.0         &187.2(11.4)& 0.53  & 0.26 & 77.5  \\
         &    &B2 &j& 0.972(0.052)& 95.6(2.4)   & 42.0(3.0) &$<0.05$&$<0.05$&      \\
         &    &A2 &j&12.773(0.079)&115.9(0.2)   & 25.5(2.2) &$<0.05$&$<0.05$&      \\
         &    &A1 &h&14.127(0.032)&115.5(0.1)   &130.4(8.0) & 0.81  & 0.23 &$-$62.1\\
1995.535 &8.4 &B1 &h& 0.0         & 0.0         &238.3(14.5)& 0.43  & 0.43 &       \\
         &    &B2 &j& 0.901(0.048)&91.5(2.4)    & 69.0(4.7) &$<0.05$&$<0.05$&      \\
         &    &A2 &j&12.893(0.086)&116.2(0.2)   & 35.4(3.1) &$<0.05$&$<0.05$&      \\
         &    &A1 &h&14.114(0.039)&115.5(0.1)   &166.4(10.4)& 0.76  & 0.57 &$-$58.3\\
\hline
\multicolumn{10}{c}{J1756+5748 (Core-Jet)} \\
2009.593 & 8.4& A1&j&12.115(0.018)&79.3(0.1)    & 48.9(4.9) & 0.65  & 0.65 &       \\
            & & A2&j&11.434(0.016)&79.4(0.1)    &123.8(12.4)& 0.52  & 0.29 & 73.1  \\
            & & B &j& 8.946(0.022)&81.2(0.1)    & 43.5(4.4) & 1.63  & 0.52 &$-$84.5\\
            & & C &j& 3.319(0.044)&79.6(0.6)    &  6.1(0.8) & 0.17  & 0.17 &       \\
            & & D2&j& 0.856(0.016)&77.1(0.9)    & 73.6(7.4) & 0.24  & 0.09 & 45.9  \\
            & &D1 &c& 0.0         &0.0          & 23.9(2.4) & 0.33  & 0.06 & 76.4  \\
2005.336 & 8.4& A1&j&12.870(0.037)&79.5(0.1)    &  17.9(1.3)& 1.04  & 0.28 & 76.2  \\
	    & & A2&j&11.492(0.014)&79.5(0.1)    & 105.7(6.4)& 0.70  & 0.29 & 85.1  \\
	    & & B &j& 8.711(0.026)&81.6(0.1)    &  42.2(2.6)& 1.75  & 0.48 &$-$83.0\\
	    & & C &j& 3.306(0.092)&77.9(1.3)    &   5.0(0.7)&$<$0.60&$<$0.60&      \\
	    & & D2&j& 0.746(0.016)&81.0(1.0)    &  49.8(3.0)&$<$0.20&$<$0.20&      \\
            & &D1 &c& 0.0         &0.0          &  19.0(1.3)&$<$0.30&$<$0.30&      \\
1998.122 & 5.0& A1&j&12.537(0.022)& 79.4(0.1)   &  57.2(3.5)& 0.94 & 0.94 & 0.0    \\
	    & & A2&j&11.071(0.017)& 79.6(0.1)   &191.7(11.5)& 0.93 & 0.32 & 78.8   \\
            & & B &j& 7.992(0.021)& 81.5(0.4)   &  79.7(4.8)& 1.67 & 0.68 &$-$81.0 \\
            & & C &j& 2.906(0.135)& 77.2(2.4)   &   8.3(0.8)& 1.74 & 1.74 &$-$161.7\\
            & &D1+D2&c& 0.0         & 0.0         &  63.4(3.8)& 0.73 & 0.12 &80.9    \\
1994.609 & 8.3& A2&j&11.203(0.085)& 79.8(0.4)   & 131.2(8.3)& 0.39 & 0.16 &$-$62.4\\
            & & B &j& 8.127(0.062)& 82.1(0.4)   &  39.1(3.6)& 2.26 & 0.20 & 1.6   \\
            & &D1+D2&c& 0.0         & 0.0         &  67.2(5.0)& 1.01 & 0.62 & 88.4  \\
\hline
\multicolumn{10}{c}{J2203+1007 (CSO)} \\
2005.336 &8.4 & A1&h& 0.0         & 0.0         &109.3(6.6) & 0.73  & 0.48 &$-$21.4\\
         &    & A2&j& 0.892(0.033)&$-$81.0(1.4) &28.8(1.8)  & 0.30  & 0.30 &       \\
         &    & B &j& 2.940(0.081)&$-$75.4(1.2) &10.5(1.0)  & 1.07  & 1.07 &       \\
         &    & D2&j& 9.659(0.045)&$-$68.6(0.2) &13.0(0.9)  & 0.00  & 0.00 &       \\
         &    & D1&h&10.447(0.032)&$-$69.9(0.1) &36.0(2.2)  & 0.61  & 0.61 &       \\
2002.919 & 8.4& A1&h& 0.0         & 0.0         &125.3(7.5) & 0.98  & 0.56 &$-$20.2\\
         &    & A2&j& 1.015(0.043)&$-$80.4(1.8) & 24.7(1.6) & 0.30  & 0.30 &       \\
         &    & B &j& 2.934(0.087)&$-$75.8(1.3) & 10.5(0.9) & 0.83  & 0.83 &       \\
         &    & D2&j& 9.437(0.035)&$-$69.3(0.1) & 13.9(1.0) & 0.0   & 0.0  &       \\
         &    & D1&h&10.402(0.061)&$-$69.7(0.2) & 40.8(2.5) & 0.61  & 0.61 &       \\
2001.010 &15.4& A1&h& 0.0         & 0.0         & 71.4(4.3) & 0.78  & 0.52 &$-$18.4\\
         &    & A2&j& 0.807(0.042)&$-$81.9(1.9) & 15.3(1.0) & 0.30  & 0.30 &       \\
         &    & B &j& 2.954(0.082)&$-$75.8(1.2) & 10.5(1.2) & 1.55  & 1.55 &       \\
         &    & D2&j& 9.795(0.071)&$-$66.6(0.3) &  4.8(0.5) & 0.0   & 0.0  &       \\
         &    & D1&h&10.436(0.047)&$-$69.9(0.1) & 17.4(1.2) & 0.83  & 0.83 &       \\
2000.230 &8.4 & A1&h& 0.0         & 0.0         &119.5(7.2) & 0.80  & 0.51 &$-$19.1\\
         &    & A2&j& 0.890(0.039)&$-$81.4(1.6) & 25.1(1.6) & 0.30  & 0.30 &       \\
         &    & B &j& 2.948(0.077)&$-$75.9(1.1) & 10.5(1.0) & 1.07  & 1.07 &       \\
         &    & D2&j& 9.606(0.045)&$-$69.0(0.1) & 15.1(1.0) & 0.0   & 0.0  &       \\
         &    & D1&h&10.429(0.037)&$-$69.9(0.1) & 36.3(2.2) & 0.58  & 0.58 &       \\
1998.204 &8.4 & A1&h& 0.0         & 0.0         & 99.6(6.0) & 0.63  & 0.41 &$-$11.0\\
         &    & A2&j& 0.807(0.052)&$-$83.1(2.7) & 19.9(1.4) & 0.10  & 0.10 &       \\
         &    & B &j& 2.872(0.123)&$-$76.7(1.7) & 10.5(1.7) & 1.23  & 1.23 &       \\
         &    & D2&j& 9.559(0.079) &$-$68.9(0.3)& 12.2(1.1) & 0.0   & 0.0  &       \\
         &    & D1&h&10.458(0.045)&$-$69.8(0.1) & 30.8(2.1) & 0.55  & 0.55 &       \\
1995.535 &8.4 & A1&h& 0.0         & 0.0         &181.4(11.1)& 0.49  & 0.49 &       \\
         &    & D2&j& 9.189(0.106)&$-$70.3(2.6) & 12.9(2.2) & 0.0   & 0.0  &       \\
         &    & D1&h&10.097(0.106)&$-$69.8(0.4) & 36.2(3.6) & 0.92  & 0.92 &       \\
\hline
\multicolumn{10}{c}{J2312+3847 (Core-Jet)} \\
2009.593 & 8.4& A &j& 6.689(0.017)& 53.9(0.2)   & 61.6(3.9) & 0.95  & 0.65  & 63.3 \\
            & & B &j& 3.054(0.024)& 47.0(0.5)   & 21.6(1.5) &$<$0.15&$<$0.15&      \\
            & & C &j& 1.874(0.015)& 38.0(0.5)   &296.9(17.8)& 0.79  & 0.22  & 48.2 \\
            & & D &c& 0.0         & 0.0         &108.2(6.5) & 0.79  & 0.14  & 37.5 \\
2005.336 & 8.4& A &j& 6.579(0.019)& 54.1(0.2)   &  66.9(4.5)& 1.41  & 0.82  & 71.5 \\
            & & B &j& 2.845(0.018)& 47.6(0.4)   &  25.1(1.8)&$<$0.15&$<$0.15&      \\
            & & C &j& 1.806(0.021)& 37.4(0.6)   &220.4(13.3)& 0.67  & 0.15  & 55.7 \\
            & & D &c& 0.0         & 0.0         &  78.3(4.8)& 0.47  & 0.47  & 40.0 \\
1996.435 & 8.3& A &j& 6.463(0.060)& 53.8(0.5)   &  87.9(5.4)& 1.42  & 0.50  & 64.1 \\
            & & B &j& 2.558(0.059)& 42.1(1.3)   &  63.0(3.9)& 0.15  & 0.15  &      \\
            & & C &j& 1.694(0.061)& 37.2(2.0)   & 129.4(7.9)& 0.80  & 0.80  &      \\
            & & D &c& 0.0         & 0.0         &  63.3(3.9)& 0.35  & 0.35  &      \\      
\label{tab:modfit}
\end{longtable} 
In column (4) the abbreviation j stands for jet, c for core, and h for hot spot.
Note-- (a) The model fitting results in epoch 1993 are from Henstock et al. (1995).
}

\begin{landscape}
\begin{table*}
 \renewcommand\tabcolsep{4pt}
\centering
\caption{Proper motions of Sources with known redshifts.}
\begin{tabular}{cccccccccc}
\hline
Source & Comp&$\mu_\alpha^a$ & $\mu_\delta^a$ & r & PA & $v_r^b$ & $v_t$ & $t^{c}_{age}$  &    N$_{Epochs}$\\
       &           & ($\mu$as yr$^{-1}$) & ($\mu$as yr$^{-1}$) & (pc) & (deg) & ($c$) & ($c$)  & (yr) & \\
   (1)  & (2) & (3) & (4) & (5) & (6) & (7) & (8) & (9) & (10) \\
\hline
J0017+5312 &  B-A(j-h) &  11.7        &$-4.0$        & 14.1 &$-$65.0 &$-$1.1         &               &              &2(1994-2005)\\
           &  C-A(h-h) &  26.4        &$-24.2$       & 40.2 &$-$65.0 &$-$3.2         & 1.0           &              &\\
J1324+4048 &  B-A(h-h) &$-4.2\pm1.4$  &$-1.0\pm1.5$  & 32.5 &$-$83.0 & 0.12$\pm$0.04 &               & 870$\pm$290  &3(1996-2009)\\
J1335+5844 &B1-A1(h-h) & $14.3\pm2.9$ &$-8.8\pm3.0$  &84.6  &$-$164.7& 0.16$\pm$0.10 & 0.53$\pm$0.10 & 1800$\pm$1150&5(1994-2006)\\
           &B2-A1(h-h) & $23.0\pm3.9$ &$-22.9\pm4.1$ &84.2  &$-$160.6& 0.45$\pm$0.13 & 0.98$\pm$0.10 & 600$\pm$170  &5(1994-2006)\\
           &B3-A1(j-h) & $53.0\pm15.8$&$-96.8\pm17.5$&75.3  &$-$170.0& 2.86$\pm$0.58 & 2.29$\pm$0.53 &  90$\pm$20   &4(2002-2006)\\
J1511+0518 &  A-B(h-c) & $7.0\pm3.4$  &              & 3.4  &   89.4 & 0.04$\pm$0.02 &               & 300$\pm$140  &6(2005-2009)\\
           & C1-B(h-c) &              & $12.0\pm3.3$ & 4.0  &$-$86.8 &               &$-0.07\pm0.02$ &              &6(2005-2009)\\
           & C2-B(h-c) &$-26.4\pm3.6$ &$-16.9\pm4.9$ & 5.3  &$-$92.3 & 0.15$\pm$0.02 & 0.09$\pm$0.03 &              &6(2005-2009)\\
J1734+0926 &A1-B1(h-h) & $7.0\pm2.9$  &              &100.6 &  115.4 & 0.25$\pm$0.12 & 0.12$\pm$0.05 & 1300$\pm$620 &5(1994-2005)\\
J1756+5748 &A2-D1(j-c) & $30.6\pm1.3$ & $9.8\pm1.3$  & 93.8 &   80.0 & 2.64$\pm$0.11 & 0.36$\pm$0.11 & 116$\pm$5    &4(1994-2009)\\ 
           & B-D1(j-c) & $80.9\pm1.8$ & $16.5\pm1.9$ & 73.4 &   80.0 & 6.85$\pm$0.15 &               &  35$\pm$1    &4(1994-2009)\\
           & C-D1(j-c) & $34.5\pm7.3$ &              & 27.2 &   80.0 & 2.82$\pm$0.60 &               &  31$\pm$7    &3(1998-2009)\\
           &D2-D1(j-c) & $22.8$       & $17.4$       &  7.0 &   80.0 &  2.1          &  1.1          &  $\sim$11    &2(2005-2009\\
J2203+1007 &D1-A1(h-h) &$-11.0\pm3.8$ &              & 81.8 & $-69.8$& 0.53$\pm$0.18 & 0.20$\pm$0.07 & 500$\pm$180  &5(1998-2005)\\
           &D2-A1(j-h) &$-16.3\pm4.4$ & $17.0\pm7.2$ & 75.7 & $-69.8$& 1.13$\pm$0.26 &               & 220$\pm$50   &4(1998-2005)\\ 
J2312+3847 &  A-D(j-c) & $23.1\pm4.3$ & $18.4\pm5.1$ & 54.7 &   53.9 & 2.49$\pm$0.39 &               &  71$\pm$16   &3(1996-2009)\\
           &  B-D(j-c) & $57.4\pm4.4$ & $25.3\pm5.1$ & 24.9 &   53.9 & 5.17$\pm$0.39 & $-1.13\pm0.41$&  16$\pm$1    &3(1996-2009)\\
           &  C-D(j-c) & $17.4\pm4.1$ & $15.1\pm4.6$ & 15.3 &   53.9 & 1.94$\pm$0.36 &               &  26$\pm$5    &3(1996-2009)\\
\hline
\end{tabular}\\ 
In column (2) the abbreviation j stands for jet, c for core, and h for hot spot.
Notes: (a) Positive values of $\mu_\alpha$ and $\mu_\delta$ represent motions to the East and North, respectively.
(b) $v_r$ is calculated by assuming a position angle for the jet component given in Column (6); a positive value is defined as the motion 
away from the reference point;
(c) The age in the source rest-frame.

\label{tab:pm}
\end{table*}
\end{landscape}

\begin{landscape}
\begin{table*}
 \renewcommand\tabcolsep{4pt}
\centering
\caption{Proper motions of Sources without redshifts.}
\begin{tabular}{lrcccccccc}
\hline
Source & Comp    &$\mu_\alpha^a$ & $\mu_\delta^a$ & r & PA & $\mu_r^b$ & $\mu_t$ & $t_{age}$ &N$_{Epoch}$\\
       &     & ($\mu$as yr$^{-1}$) & ($\mu$as yr$^{-1}$) & (mas) & (deg) & ($\mu$as yr$^{-1}$) & ($\mu$as yr$^{-1}$)  & (yr) & \\
   (1)  & (2) & (3) & (4) & (5) & (6) & (7) & (8) & (9) &(10) \\
\hline
J0132+5620 & B1-A(j-h) & $71.1\pm6.8$ &$-15.8\pm8.3$ & 2.383&$-$80.7 &$-72.7\pm6.8$ &              &                      &4(1998-2009)\\
           &  C-A(j-h) &$-57.7\pm4.0$ & $16.4\pm4.1$ &10.539&$-$80.7 & $59.6\pm4.0$ &              &$\frac{180\pm12}{1+z}$&4(1998-2009)\\ 
           & D2-A(j-h) &$-27.4\pm1.8$ & $5.0\pm2.2$  &11.547&$-$80.7 & $27.8\pm1.8$ &              &$\frac{410\pm25}{1+z}$&7(1994-2009)\\ 
           & D1-A(h-h) & $5.4\pm1.3$  & $2.9\pm1.3$  &12.181&$-$80.7 &$-4.9\pm1.3$  & $-3.7\pm1.3$ &$>\frac{4700}{1+z}^c$ &6(1994-2005)\\ 
J0518+4730 &  A-B(h-c) & $26.4\pm2.7$ &$-15.3\pm3.1$ & 1.921& 132.1  & $29.8\pm2.9$ & $6.3\pm2.9$  &                      &5(1996-2005)\\ 
           &  C-B(h-c) & $21.0\pm1.9$ &$-16.0\pm2.0$ & 3.054&$-$65.8 &$-25.7\pm1.9$ &  $6.0\pm2.0$ &$\sim\frac{1200}{1+z}^d$&5(1996-2005)\\
\hline
\end{tabular}\\ 
In column (2) the abbreviation j stands for jet, c for core, and h for hot spot.
Note: (a) Positive values of $\mu_\alpha$ and $\mu_\delta$ represent motions to the East and North, respectively;
(b) $\mu_r$ is calculated by assuming a mean position angle for all jet components; a positive value is defined as the motion away from the reference point;
(c) This lower limit of the kinematic age is calculated assuming the expansion velocity to be 2-$\sigma$ uncertainty, see Section 4.2;
(d) The kinematic age is calculated from the arm length and the HS-HS expansion velocity, see discussion in Section 4.3.

\label{tab:pm2}
\end{table*}
\end{landscape}

\end{document}